\documentclass{aastex63}

\usepackage{natbib}
\usepackage{threeparttable}
\usepackage{color}
\usepackage{graphicx}
\usepackage{epstopdf}
\usepackage{amsthm,amsmath,amssymb}
\usepackage{mathrsfs}

\usepackage{txfonts}

\shorttitle{Observations of SN 2012ij}
\shortauthors{Li, Z., Zhang, T. et al.}

\begin{document} 

\title{SN 2012ij: A low-luminosity type Ia supernova and evidence for continuous distribution from 91bg-like explosion to normal ones \footnote{This paper includes data gathered with the 6.5 meter Magellan Telescopes located at Las Campanas Observatory, Chile.}}

\author{Zhitong Li}
\affil{Key Laboratory of Optical Astronomy, National Astronomical Observatories, Chinese Academy of Sciences, Beijing 100101, China}
\affil{School of Astronomy and Space Science, University of Chinese Academy of Sciences, Beijing 101408, China}

\author{Tianmeng Zhang}
\email{zhangtm@nao.cas.cn}
\affil{Key Laboratory of Optical Astronomy, National Astronomical Observatories, Chinese Academy of Sciences, Beijing 100101, China}
\affil{School of Astronomy and Space Science, University of Chinese Academy of Sciences, Beijing 101408, China}

\author{Xiaofeng Wang}
\email{wang\_xf@mail.tsinghua.edu.cn}
\affil{Physics Department and Tsinghua Center for Astrophysics (THCA), Tsinghua University, Beijing, 100084, China}
\affil{Beijing Planetarium, Beijing Academy of Science and Technology, Beijing, 100044, China}
\author{Hanna Sai}
\affil{Physics Department and Tsinghua Center for Astrophysics (THCA), Tsinghua University, Beijing, 100084, China}
\author{Jujia Zhang}
\affil{Yunnan Observatories, Chinese Academy of Sciences, Kunming 650216, China}
\affil{Key Laboratory for the Structure and Evolution of Celestial Objects, Chinese Academy of Sciences, Kunming 650216, China}
\author{Juncheng Chen}
\affil{Wu Zhou University, Wuzhou 543002, China}
\author{Xulin Zhao}
\affil{Department of Physics, Tianjin University of Technology, Tianjin 300384, China}
\author{Shengyu Yan}
\affil{Physics Department and Tsinghua Center for Astrophysics (THCA), Tsinghua University, Beijing, 100084, China}
\author{Bo Wang}
\affil{Yunnan Observatories, Chinese Academy of Sciences, Kunming 650216, China}
\affil{Key Laboratory for the Structure and Evolution of Celestial Objects, Chinese Academy of Sciences, Kunming 650216, China}

\author{Mark M. Phillips}
\affil{Carnegie Observatories, Las Campanas Observatory, Casilla 601, La Serena, Chile}
\author{Eric Y. Hsiao}
\affil{Carnegie Observatories, Las Campanas Observatory, Casilla 601, La Serena, Chile}
\author{Nidia Morrell}
\affil{Carnegie Observatories, Las Campanas Observatory, Casilla 601, La Serena, Chile}
\author{Carlos Contreras}
\affil{Carnegie Observatories, Las Campanas Observatory, Casilla 601, La Serena, Chile}
\author{Christopher R. Burns}
\affil{Observatories of the Carnegie Institution for Science, 813 Santa Barbara Street, Pasadena, CA 91101, USA}
\author{Christopher Ashall}
\affil{Institute for Astronomy, University of Hawai'i at Manoa, 2680 Woodlawn Dr. Hawai'i, HI 96822, USA}
\author{Maximilian Stritzinger}
\affil{Department of Physics and Astronomy, Aarhus University, Ny Munkegade 120, DK-8000 Aarhus C, Denmark}
\author{Kevin Krisciunas}
\affil{George P. and Cynthia Woods Mitchell Institute for Fundamental Physics and Astronomy, Texas A\&M University, Department of Physics and Astronomy, College Station, TX 77843, USA}
\author{Jose Prieto}
\affil{Department of Astrophysical Sciences, Princeton University, 4 Ivy Ln., Princeton, NJ 08544, USA}

\author{Hu Zou}
\affil{Key Laboratory of Optical Astronomy, National Astronomical Observatories, Chinese Academy of Sciences, Beijing 100101, China}
\author{Jiali Wang}
\affil{Key Laboratory of Optical Astronomy, National Astronomical Observatories, Chinese Academy of Sciences, Beijing 100101, China}
\author{Jun Ma}
\affil{Key Laboratory of Optical Astronomy, National Astronomical Observatories, Chinese Academy of Sciences, Beijing 100101, China}
\affil{School of Astronomy and Space Science, University of Chinese Academy of Sciences, Beijing 101408, China}
\author{Jundan Nie}
\affil{Key Laboratory of Optical Astronomy, National Astronomical Observatories, Chinese Academy of Sciences, Beijing 100101, China}
\author{Suijian Xue}
\affil{Key Laboratory of Optical Astronomy, National Astronomical Observatories, Chinese Academy of Sciences, Beijing 100101, China}
\author{Xu Zhou}
\affil{Key Laboratory of Optical Astronomy, National Astronomical Observatories, Chinese Academy of Sciences, Beijing 100101, China}
\author{Zhimin Zhou}
\affil{Key Laboratory of Optical Astronomy, National Astronomical Observatories, Chinese Academy of Sciences, Beijing 100101, China}
\author{Danfeng Xiang}
\affil{Physics Department and Tsinghua Center for Astrophysics (THCA), Tsinghua University, Beijing, 100084, China}
\author{Gaobo Xi}
\affil{Physics Department and Tsinghua Center for Astrophysics (THCA), Tsinghua University, Beijing, 100084, China}

\begin{abstract}

In this paper, we present photometric and spectroscopic observations of a subluminous type Ia supernova (SN Ia) 2012ij, which has an absolute $B$-band peak magnitude $M_{B,\rm{max}}$ = $-$17.95 $\pm$ 0.15 mag. The $B$-band light curve exhibits a fast post-peak decline with $\Delta m_{15}(B)$ = 1.86 $\pm$ 0.05 mag. All the $R$ and $I$/$i$-band light curves show a weak secondary peak/shoulder feature at about 3 weeks after the peak, like some transitional subclass of SNe Ia, which could result from an incomplete merger of near-infrared (NIR) double peaks. The spectra are characterized by Ti~{\sc ii} and strong Si~{\sc ii} $\lambda$5972 absorption features that are usually seen in low-luminosity objects like SN 1999by. The NIR spectrum before maximum light reveals weak carbon absorption features, implying the existence of unburned materials. We compare the observed properties of SN 2012ij with those predicted by the sub-Chandrasekhar-mass and the Chandrasekhar-mass delayed-detonation models, and find that both optical and NIR spectral properties can be explained to some extent by these two models. By comparing the secondary maximum features in $I$ and $i$ bands, we suggest that SN 2012ij is a transitional object linking normal SNe Ia to typical 91bg-like ones. From the published sample of SNe Ia from the $Carnegie~Supernova~Project~II$ (CSP-II), we estimate that the fraction of SN 2012ij-like SNe Ia is not lower than $\sim$ 2\%.

\end{abstract}


\section{Introduction}\label{sec:intro}

Owing to high luminosities and relatively uniform light curves, SNe Ia have been used to estimate the Hubble constant (e.g., \citealt{ham96a,sand06,riess07,riess16}) and to determine the expansion history of the universe \citep{riess98,per99,bet14}. Some empirical relations found between their peak luminosities and light/color-curve shapes (e.g., \citealt{phi93,guy05,wxf05,wxf06,jha07}) make them good distance indicators. It is widely accepted that SNe Ia arise from thermonuclear explosions of carbon-oxygen white dwarfs (C-O WDs) in close binary systems. Two popular scenarios have been proposed for the companion star of the exploding WD. Candidate companion stars could be a non-degenerate object (i.e., main sequence, red-giant, subgiant star, or helium star), dubbed as single degenerate (SD) system \citep{whe73}, or another WD, dubbed as the double degenerate (DD) scenario \citep{iben84, web84}. Both SD \citep{pat07, ster11, dil12, cao15, hoss17, dimi19, shappee19, liw19a} and DD \citep{liw11b, schae12, edw12, sant15} systems are supported by some observations, and it seems that both models are required to explain the observed diversity among SNe Ia \citep{wb12, wxf13b, maoz14, wb18}.

Most of SNe Ia ($\sim$70\%) display relatively uniform properties in photometric and spectroscopic evolution, which have been called Branch-normal ones \citep{bran93, fili97}. In contrast to the Branch-normal SNe Ia, some SNe Ia are classified into different sub-classes due to different photometric or spectroscopic evolution. The over-luminous group like 91T SNe Ia are characterized by broad light curves, relatively weak Si~{\sc ii} $\lambda$6355~{\AA} and prominent Fe~{\sc ii/iii} absorption features around the maximum light \citep{fili92a, phi92, ruiz92}. On the contrary, the sub-luminous SNe Ia like SN 1991bg showed fast-evolving light curves, prominent absorptions of intermediate-mass elements \citep{fili92b,lei93}. The mass of of $^{56}$Ni created in explosion of 91bg-like SNe was much lower than that of normal events \citep{maz97,maoz14}. Due to their low luminosities, the 91bg-like subclass of SNe Ia was only 3\% in a magnitude-limited SNe Ia sample \citep{liw11a}. Transitional objects like SNe 1986G\citep{hou87,phi87,phi99} and 2011iv\citep{gal18} show photometric and spectroscopic properties inbetween the normal and 91bg-like sub-classes of SNe Ia, which may appear at least as frequent as the 91bg-like objects \citep{gon11}, suggested a continuous distribution from normal to 91bg-like SNe Ia. There is another subclass of low luminosity SNe Ia like SN 2002es are similar to 91bg-like and 86G-like SNe Ia which share some common spectroscopic features as the 91bg-like objects but it has much wider light curves in comparison \citep{tau17}. Several classification schemes, based on the velocity evolution (or gradient) of Si~{\sc ii} $\lambda$6355 \citep{ben05}, the pseudo-Equivalent Width (pEW) of Si~{\sc ii} $\lambda$6355 and Si~{\sc ii} $\lambda$5972 absorptions \citep{bran06,bur20} or the Si~{\sc ii} velocity measured at around the maximum light \citep{wxf09a}, have been proposed for SNe Ia. The above three classifications partially overlap with each other. For example, the cool subclass \citep{bran06}, characterized by low temperature and strong Si~{\sc ii} $\lambda$5972 absorption, has a significant overlap with the 91bg-like subclass. 

Some observational characteristics were used to distinguish 91bg-like SNe. Photometrically, 91bg-like SNe Ia lack prominent secondary maximum features in near-infrared (NIR) bands, likely due to an early recombination of iron-group elements (IGE) \citep{tau17}. Spectroscopically, 91bg-like SNe display strong Si~{\sc ii} $\lambda$5972 and prominent Ti~{\sc ii} lines, suggestive of low photospheric temperatures \citep{nug95}. The observed differences between 91bg-like SNe Ia and the normal ones could be attributed to different progenitor scenarios and/or explosion channels \citep{tau17,pol19}. Some studies favored the sub-$M_{\rm{Ch}}$ explosion model \citep{mae16}, while some preferred the delayed-detonation model \citep{hof02}. 

Due to the lack of a large well-observed sample, the progenitor properties and explosion mechanisms of sub-luminous 91bg-like SNe Ia still remain controversial. One of the well-observed sample is SN 1999by, for which analysis of the NIR spectra suggests that the explosion is consistent with the delayed detonation model \citep{hof02}. Polarization observations suggested that the explosion of SN 1999by had an axis of symmetry, which could be explained by rapid rotation of the progenitor WD or an explosion during the merger process of double WDs \citep{how01}. Moreover, \citet{blond18} also used a sub-Chandrasekhar-mass (sub-$M_{\rm{Ch}}$) model of pure central detonation of a C-O WD to successfully reproduce the observed properties of SN 1999by. 

One probable explosion model for 91bg-like SNe Ia is the violent merger of WD binary. In this scenario, the lighter companion is disrupted during the merging process and accreted by the massive WD. During the merging process, the accretion stream hits the surface of the WD, causing a high temperature that triggers carbon detonation \citep{pak10,pak12}. This scenario should be recognized as a sub-$M_{\rm{Ch}}$ explosion according to the density of the primary WD when it is ignited \citep{mae16}, though the total mass exceeds the Chandrasekhar mass \citep{sat15}. 

Another possible channel is the double detonation model in a binary system where the companion is a helium star or contains some helium. The primary C-O WD accretes the helium from the companion and ignites the helium shell, then the core explosion is triggered by the helium detonation on the surface. This channel does not require a Chandrasekhar mass for the exploding WD \citep{fin10}. It is also called as sub-$M_{\rm{Ch}}$ helium detonation model. As a sub-$M_{\rm{Ch}}$ explosion, the results predicted by double detonation model are consistent with the observations of sub-luminous SNe Ia with fast-evolving light curves \citep{woo94b,hoe96,ruiz93}.

Besides the sub-$M_{\rm{Ch}}$ explosions, the delayed-detonation model is also a promising scenario. In this scenario, the initial deflagration transforms into a supersonic detonation at some point, when the explosion can produce suitable nucleosynthesis for most SNe Ia \citep{kho91,woo90,hoe96,iwa99,hof02,hoe17}. 

In this paper, we present photometric and spectroscopic observations of the subluminous, peculiar type Ia SN 2012ij. Observations and data reductions are described in section \ref{observation}. In section \ref{photometricProperty}, we present the light/color curve evolution as well as the reddening estimation. Section \ref{sectionSpectra} describes the spectroscopic evolution in both optical and NIR bands. The origin of weak NIR secondary peak and the comparison with different models are discussed in section \ref{discussion}. We summarize in section \ref{conclusion}.

\section{Observations and Data reduction}\label{observation}

SN 2012ij was discovered on December 29.78 UT 2012 by the 0.6-m Schmidt telescope in the course of the THU-NAOC\footnote{National Astronomical Observatories, Chinese Academy of Sciences} Transient Survey (TNTS; \citealt{zhangtm15}), with an unfiltered magnitude of about 18.0 mag. The coordinates of this object are $\alpha$ = 11$^{\rm{h}}$40$^{\rm{m}}$15$^{\rm{s}}$.84, $\delta$ = +17$^{\circ}$27$\arcmin$22$\arcsec$.2 (J2000.0), located 4$\arcsec$.0 west and 4$\arcsec$.0 south of the center of an S0/a galaxy CGCG 097-050 \citep{hak16}.
The spectrum of the host galaxy from Sloan Digital Sky Survey (SDSS) shows no significant H$\alpha$ emission. Fitting the spectrum with Firefly \citep{wil17}, we obtained the stellar mass as $M_{*}~ \sim ~1.6 \times 10^{9}~M_{\odot}$, the age as 3.73 Gyr, and the metallicity of the stellar populations as [Z/H] = 0.16, respectively. With the passive model for passive galaxies \citet{mar09}, the model fitting gave the stellar mass as $M_{*}~ \sim ~1 \times 10^{10}~M_{\odot}$ and the age as 4.25 Gyr, respectively. Both model fittings indicate that the host galaxy of SN 2012ij is a passive galaxy, which is similar to host galaxies of most 91bg-like SNe Ia \citep{liw11a}.

\begin{figure}[ht]
\includegraphics[angle=0,width=140mm]{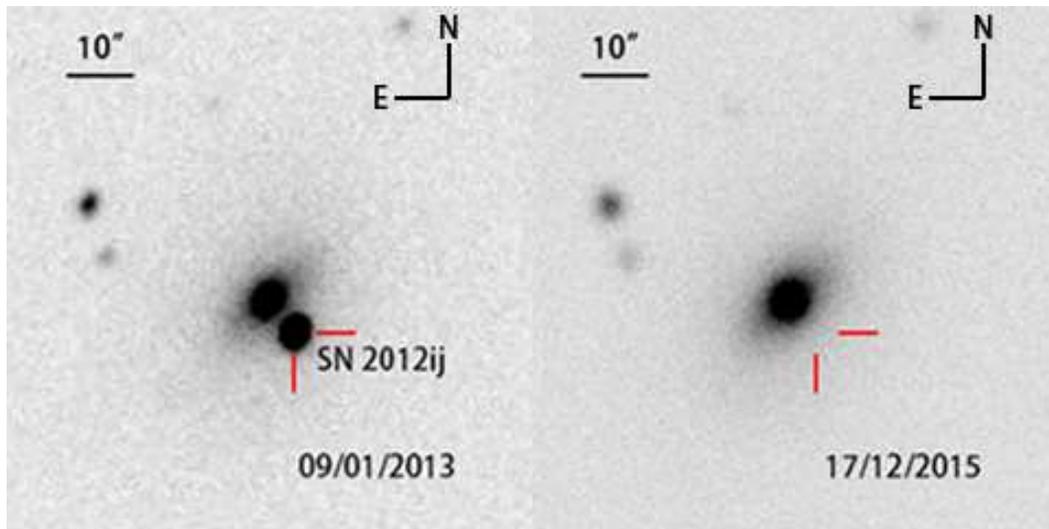}
\centering
\caption {\textit{Left}: the unfiltered image of SN 2012ij (marked by the crosshair) taken by TNTS on Jan. 09th, 2013. \textit{Right}: the unfiltered image of the same area taken by TNTS on Dec. 17th, 2015 when SN 2012ij faded away. The position of SN is marked by the crosshair.}
\label{findingchart}
\end{figure}

The finding chart of this SN is shown in Figure \ref{findingchart}. The redshift of the host galaxy is $z$ = 0.011 \citep{lu93}, corresponding to a distance modulus of $m$ $-$ $M$ = 33.54 $\pm$ 0.15 mag \footnote{Given by NASA/IPAC Extragalactic Database \citep{NEDGWF}, assuming the cosmology parameters $\Omega_{0}$ = 0.308 and H$_0$ = 67.8 $km~s^{-1}~Mpc^{-1}$ (Virgo Infall only).}. A spectrum taken at a few days after the discovery revealed that SN 2012ij was a sub-luminous type Ia SN at about one week before the maximum light \citep{wxf13a}, with Si II velocity being about 11,400 km s$^{-1}$ inferred from  Si~{\sc ii} $\lambda$6355 absorption feature.

\subsection{Photometry}

The optical photometry ($BVRI$) of SN 2012ij was mainly collected by the 0.8-m Tsinghua-NAOC Telescope (TNT\footnote{This telescope is co-operated by Tsinghua University and NAOC.}) located at Xinglong Observatory of NAOC. This telescope is equipped with a 1340 $\times$ 1300 pixel back-illuminated CCD, with a field of view (FoV) of 11.5$\arcmin$ $\times$ 11.2$\arcmin$ (pixel size $\sim$ 0.52$\arcsec$ pixel$^{-1}$) \citep{wxf08,huang12}. The TNT instrumental magnitudes were obtained using an ad hoc pipeline (based on the IRAF \footnote{IRAF, the Image Reduction and Analysis Facility, is distributed by the National Optical Astronomy Observatories, which are operated by the Association of Universities for Research in Astronomy (AURA), Inc., under cooperative agreement with the National Science Foundation.} DAOPHOT package; \citealt{stet87}). The instrumental magnitudes of TNT were calibrated by the \textbf{Sloan} catalogues converted to $BVRI$ with the relation proposed by \citet{cho08}. SN 2012ij were also monitored in $BVugri$ bands by the CSP-II \citep{phi19,hsi19}) with the Swope 1-m telescope at the Las Campanas Observatory (LCO). The data was reduced and converted into standard system following that described in \citet{kri17} and \citet{phi19}. The final flux-calibrated magnitudes of SN 2012ij are presented in Table \ref{table12ij}.

As the flux contamination caused by the host galaxy could not be neglected, the template images of SN 2012ij, taken at almost 3 years after the outburst, were used to perform the image subtraction before we conducted the photometry.

\subsection{Spectroscopy}

After the discovery, extensive optical and NIR follow-up observations were conducted for SN 2012ij on several facilities. A total of 14 spectra were obtained for SN 2012ij, including 4 optical spectra from the Xinglong 2.16-m Telescope (XLT; \citealt{XLT216}), 3 optical spectra from Lijiang 2.4-m Telescope(LJT; \citealt{LJT,YFOSC}), 4 optical spectra from LCO du Pont and Nordic Optical Telescope (NOT), and 3 NIR spectra from Folded-port Infrared Echellette (FIRE) mounted on the \textit{Magellan Baade} Telescope (CSP-II). A journal of spectroscopic observations is listed in Table \ref{table12ijspec}.

We reduced our optical spectra using the standard IRAF routines, including corrections for bias, flat field, and removal of cosmic rays. The wavelength of the spectra was calibrated by arc lamp spectra, and the flux was calibrated by standard star, which were obtained on the same night. For all of the spectra, the atmospheric extinction was also corrected using mean extinction curves obtained at the corresponding observatories. The NIR spectra were reduced using the IDL pipeline \textbf{firehose} \citep{sim13}, which was specifically designed for the reduction of FIRE data and was described in \citet{hsi19}.

\section{Photometric Properties} \label{photometricProperty}

\subsection{Optical Light Curves} \label{sec_OpticalLightCurves}

\begin{figure}[ht]
\centering
\includegraphics[angle=0,width=140mm]{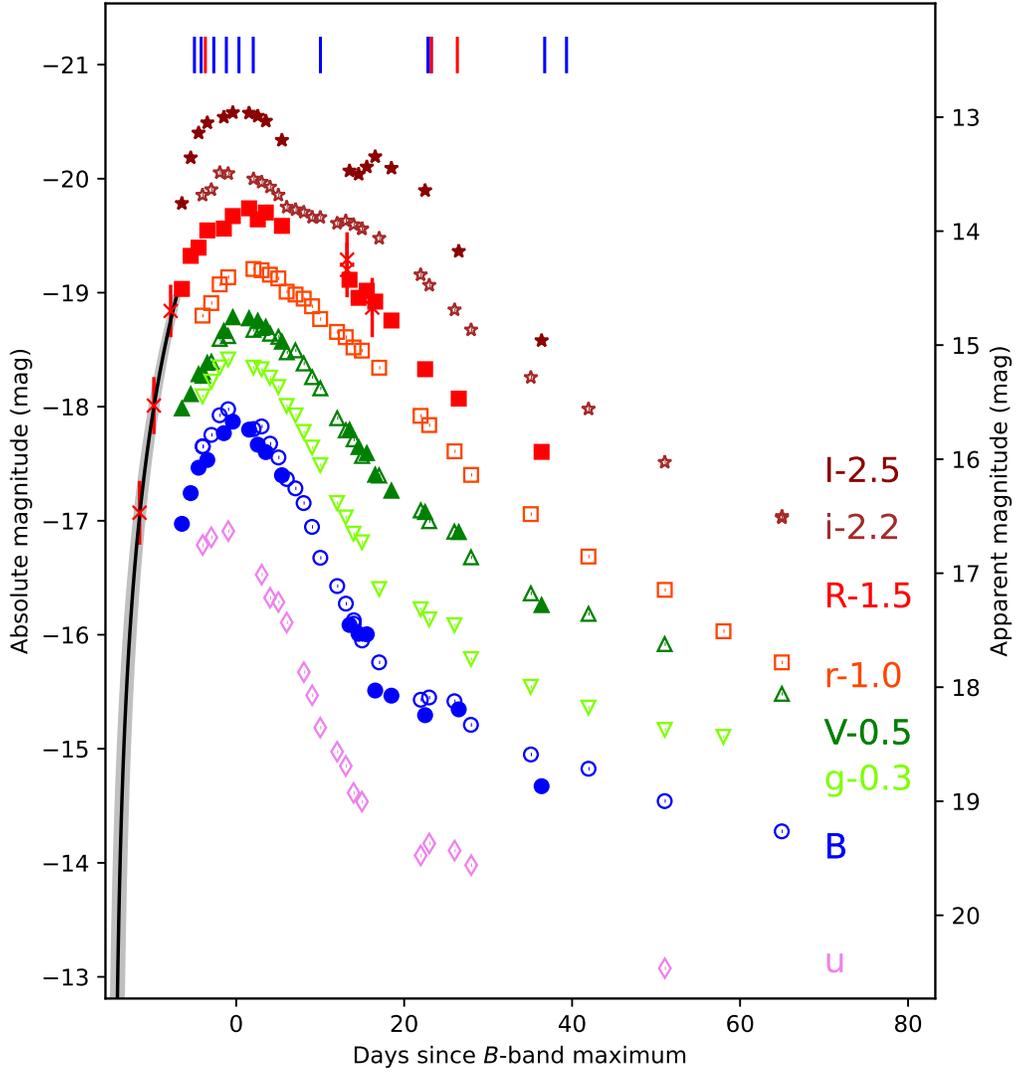}
\caption{$BVRIugri$ light curves of SN 2012ij obtained by TNT (solid markers) and CSP-II (empty markers). Unfiltered data calibrated by the $R$-band magnitude are shown with red cross. The solid line shows the early-time evolution estimated by unfiltered data, and the gray region represents the 3$\sigma$ uncertainty. Extinction due to Milky Way dust and host galaxy dust is ignored. The blue or red lines on the top of figure indicate the phases when optical (blue) or near-infrared (red) spectra were taken.}
\label{LC12ij}
\end{figure}

The $uBVgrRiI$ light curves of SN 2012ij are shown in Figure \ref{LC12ij}. These multi-band light curves were used to determine the epoch of maximum light as well as other important light curve parameters by template fitting. We found that SN 2012ij reached the $B$-band maximum light on MJD 56302.3 $\pm$ 0.3, with $B_{\rm{max}}$ = 15.67 $\pm$ 0.02 mag and the post-maximum decline rate ${\Delta}m_{15}(B)$ \citep{phi93} as 1.86 $\pm$ 0.05 mag, which was estimated in the rest frame of the SN. 

The unfiltered light curve of SN 2012ij obtained by the 0.6-m Schmidt telescope is also shown in Figure \ref{LC12ij}, and the corresponding magnitudes are listed in Table \ref{table12ijunf}. The unfiltered magnitudes were calibrated by the $R$-band magnitudes from the Positions and Proper Motions Star Catalogue Extended (PPMX; \citealt{ros08}), with a systematic uncertainty of about 0.2 - 0.3 mag \citep{liw03}. The explosion date was estimated as MJD 56288.2 $\pm$ 0.9 by fitting the first three unfiltered data points with the fireball (i.e., L $\sim$ t$^2$) model \citep{riess99}, which gives the rise time as 14.6 $\pm$ 0.9 days \footnote{Given by Monto-Carlo simulation.}. This rise time is comparable to that of some 91bg-like SNe Ia (i.e., 13 - 15 days; \citealt{tau08}). The main photometric parameters of SNe 2012ij are listed in Table \ref{tableof2obj}. For the subsequent discussions, the phases are referred with respect to the $B$-band maximum. 

\begin{figure}[ht]
\centering
\includegraphics[angle=0,width=180mm]{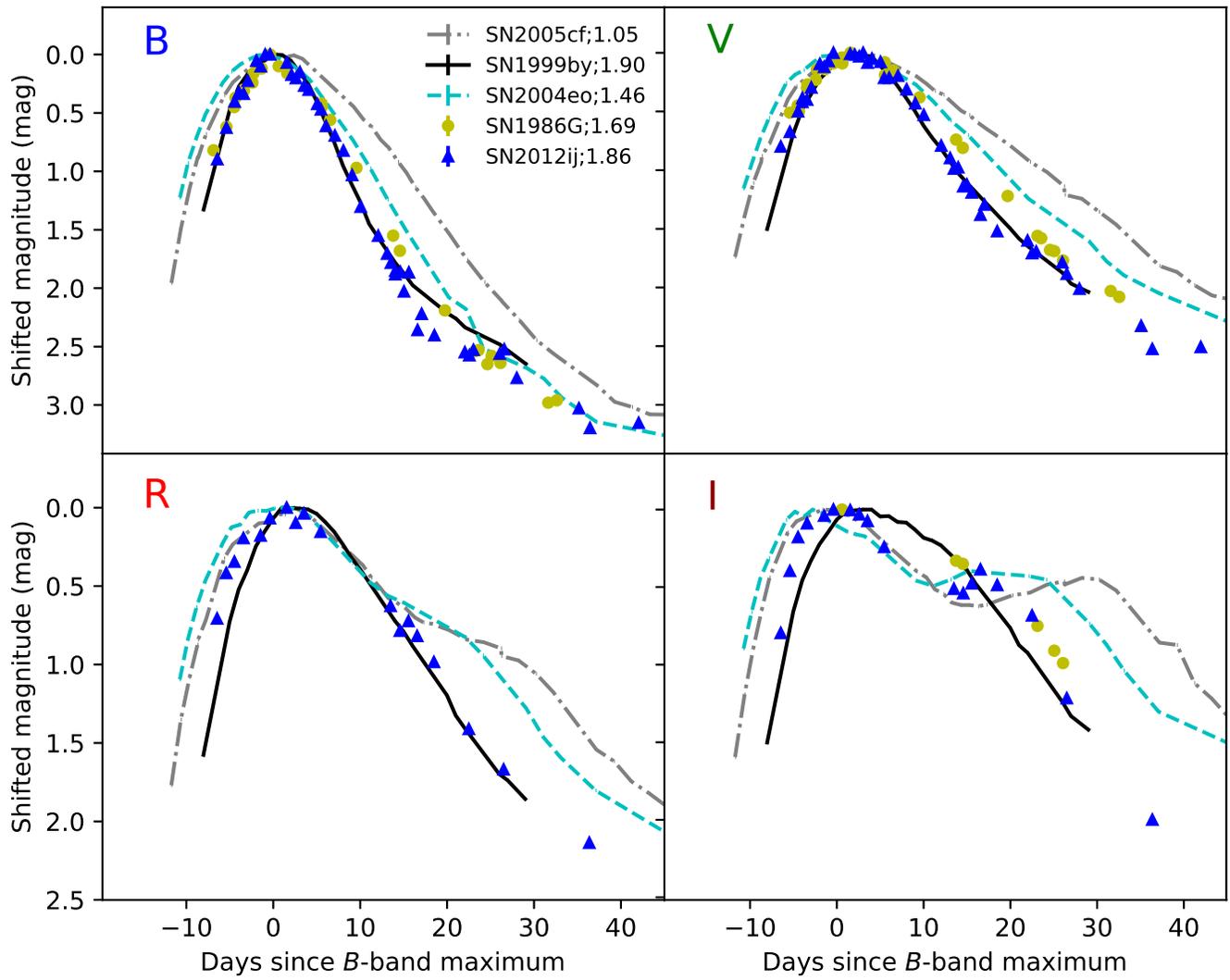}
\caption{$BVRI$ light curves of SN 2012ij from TNT, and the comparison SNe 1986G, 1999by, 2004eo and 2005cf with $\Delta m_{15}$ after their names. All light curves are shifted in magnitude to match the peak in each bands.}
\label{LC compare 1}
\end{figure}

In Figure \ref{LC compare 1}, we compare the light curves of SNe 2012ij with those of some normal and sub-luminous SNe Ia, including SNe 1986G \citep{hou87,phi87,phi99}, 1999by \citep{gar04,gane10}, 2004eo \citep{pas07b}, and 2005cf \citep{wxf09b}. One can see that the light curves of SN 2012ij are similar to those of some 91bg-like SNe in $BVR$ bands in terms of the morphology except in $I$, where SN 2012ij clearly shows post-peak shoulder/bump feature, while such a feature is usually absent in 91bg-like objects (e.g., SNe 1999by, 2006mr, see \citealt{con10}). Moreover, SN 2012ij arrived at the $I$- band primary peak $\sim$3 days earlier than SN 1999by relative to their $B$- band peaks. We will discuss the NIR features in detail in section \ref{discussion_lc}.

\subsection{The Reddening and Color Curves} \label{reddening}

The Galactic reddening toward SN 2012ij was estimated as $A_{V}^{Gal}$ = 0.073 \citep{schla11}. Since this SN locates in the outer part of the host galaxy and the SN spectra did not show obvious absorption of Na~{\sc i} D, the host-galaxy reddening was thus assumed to be negligible. Taking the Galactic reddening into account and adopting R$_V$ = 3.1 \citep{car89,odo94}, the $B$-band peak absolute magnitude of SN 2012ij was estimated as $M_{B, \rm{max}}$ = $-$17.95 $\pm$ 0.15 mag.

\begin{figure}[ht]
\centering
\includegraphics[angle=0,width=140mm]{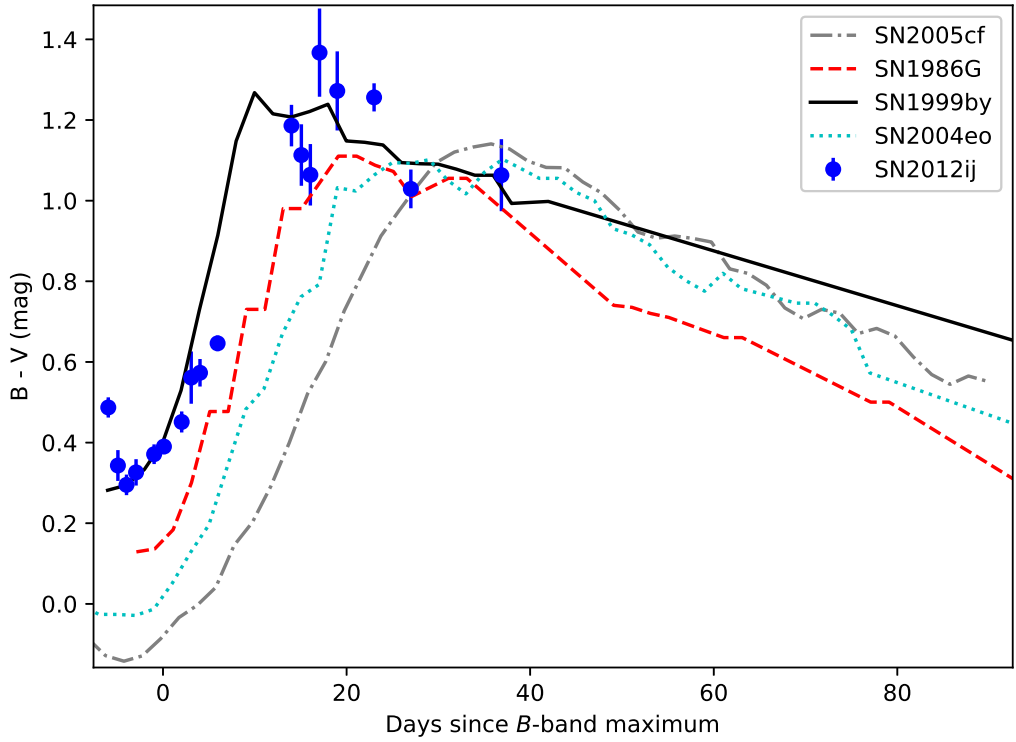}
\caption{The $B-V$ color curves of SN 2012ij, compared to standard type Ia SN 2005cf, typical 91bg-like SN 1999by, and transitional subclass of SNe 1986G and 2004eo. All SNe have been dereddened.}
\label{color}
\end{figure}

In Figure \ref{color}, we compare the $B-V$ color curve of SN 2012ij with those of 91bg-like and normal SNe Ia. One can find that SN 2012ij has systematically much redder colors than normal SNe Ia and it reached the reddest color at about 10 days earlier than the latter. After t $\sim$ 35 days from the maximum, their color curves became indistinguishable. 

The color-stretch factor, defined as $s_{BV}=t_{(B-V) \rm{max}}(days)/{30}$, is proportional to the epoch when the $B-V$ color reaches the reddest value. \citet{burns14} suggested that this factor was better than ${\Delta}m_{15}(B)$, especially when dealing with the fast-evolving 91bg-like SNe. The reason is that the $B$-band light curves of fast decliners tend to flatten earlier than t $\sim$ +15 days. Adopting the $B-V$ color curve of SN 1999by as a template, we obtained $s_{BV}$ = 0.53 for SN 2012ij.

\begin{figure}[ht]
\centering
\includegraphics[angle=0,width=180mm]{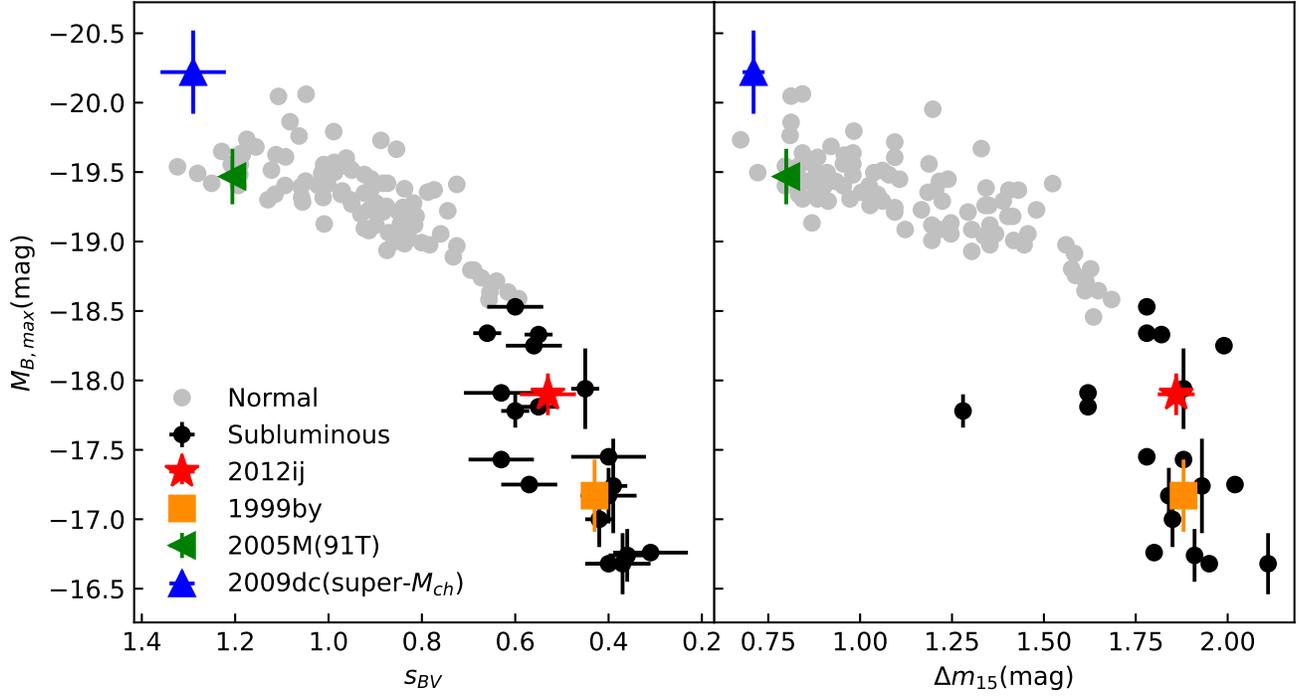}
\caption{\textit{Left}: $M_{B,\rm{max}}$ of SNe Ia are plotted against their $s_{BV}$. \textit{Right}: $M_{B,\rm{max}}$ of the same sample of SNe Ia are plotted against their post-peak decline rate $\Delta m_{15}(B)$. The sample includes SN 2012ij (red star), other subluminous SNe Ia (black points; see Table \ref{tableOthers}), 91T-like SN 2005M \citep{fre09}, super-Chandrasekhar SN 2009dc \citep{tau11}, and normal SNe Ia (\citealt{kri17,gal19}). }.
\label{sbv}
\end{figure}

A larger sample, including sub-luminous and over-luminous SNe Ia collected through $the~Open~Supernova~Catalog$ (shown in Table \ref{tableOthers}) and those released in \citet{kri17}, are used to study the distribution of color-stretch factors. In Figure \ref{sbv}, we present the correlations between their absolute $B$-band peak magnitudes, stretch factors $s_{BV}$ and decline rate ${\Delta}m_{15}(B)$. One can see that SN 2012ij locates inbetween normal SNe Ia and 91bg-like objects, favoring for a continuous distribution between normal and 91bg-like SNe Ia instead of two totally different populations. 

\subsection{Bolometric Light Curves} \label{sec_Bolometric}

\begin{figure}[ht]
\centering
\includegraphics[angle=0,width=140mm]{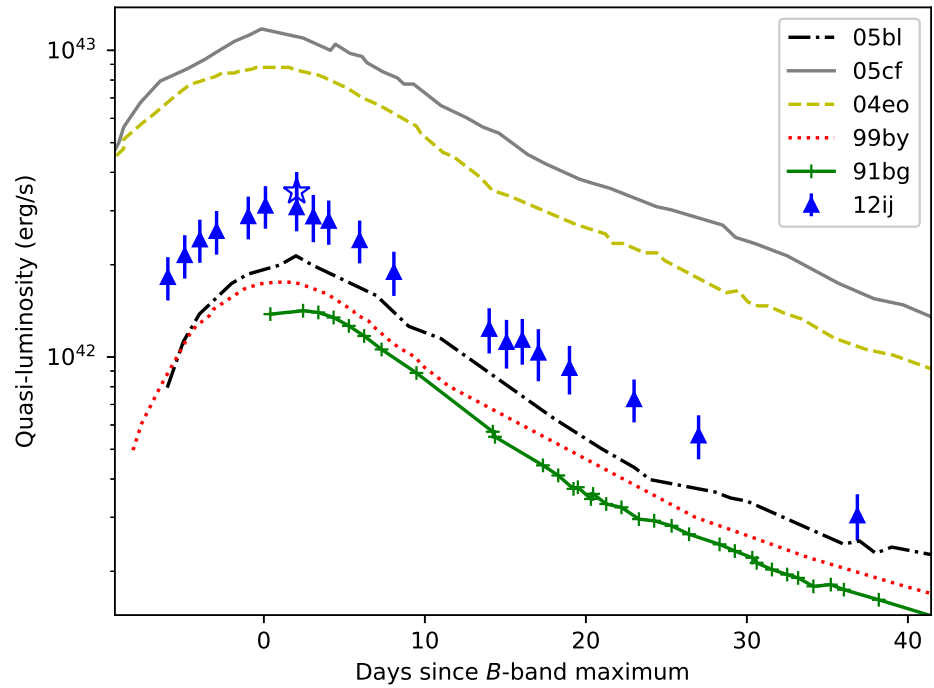}
\caption{Quasi-bolometric light curve of SN 2012ij compared with SNe 1991bg, 1999by, 2004eo, 2005cf and 2005bl in \citet{tau08}. Predicted peak real bolometric luminosities of SN 2012ij is noted with empty star.}
\label{bol}
\end{figure}

The quasi-bolometric light curves could be constructed by $uBVRI$-band photometry. The quasi-bolometric flux was derived by trapezoidal integration of flux densities in $uBVRI$ bands according to response curves of different filters, which covers the emission from 3000{\AA} to 9700{\AA}, as shown in Figure \ref{bol}. 

Comparing the quasi-bolometric light curves with those of sub-luminous SNe 1991bg \citep{lei93,tur96}, 1999by \citep{gar04}, 2005bl \citep{tau08}, a transitional SN 2004eo \citep{pas07b} and a normal SN Ia 2005cf \citep{wxf09b}, we find that SN 2012ij is quite similar to SN 2005bl. After taking into account the UV and NIR corrections, which are based on the swift UV photometry ($UVW2$, $UVM2$, $UVW1$) from `$Gehrels$ $Swift$' Optical/Ultraviolet Supernova Archive (SOUSA; \citealt{brown14}) and NIR spectra taken near the maximum light, we estimate that SN 2012ij reached its maximum luminosity at $L_{\rm{bol}}$ = 3.46 ($\pm$ 0.52) $\times$ 10$^{42}$ erg s$^{-1}$. According to the radioactive decay law put forward by \citet{arn82}, adopting the rise time t$_{\rm{r}}$ = 14.6 days, and the ratio of bolometric to radioactive luminosities $\gamma$ = 1.0 for SN 2012ij \citep{nug95,hoe96,gon12}, the nickel mass of SN 2012ij is estimated as $M(^{56}\rm{Ni})$ = 0.14 $\pm$ 0.03 $M_{\odot}$, which is close to some 91bg-like SNe Ia ($\sim$ 0.07 $M_{\odot}$; \citealt{maz97}).

\section{Spectroscopic Properties}\label{sectionSpectra}

\subsection{Evolution of the Spectra}

\begin{figure}[ht]
\centering
\includegraphics[angle=0,width=160mm]{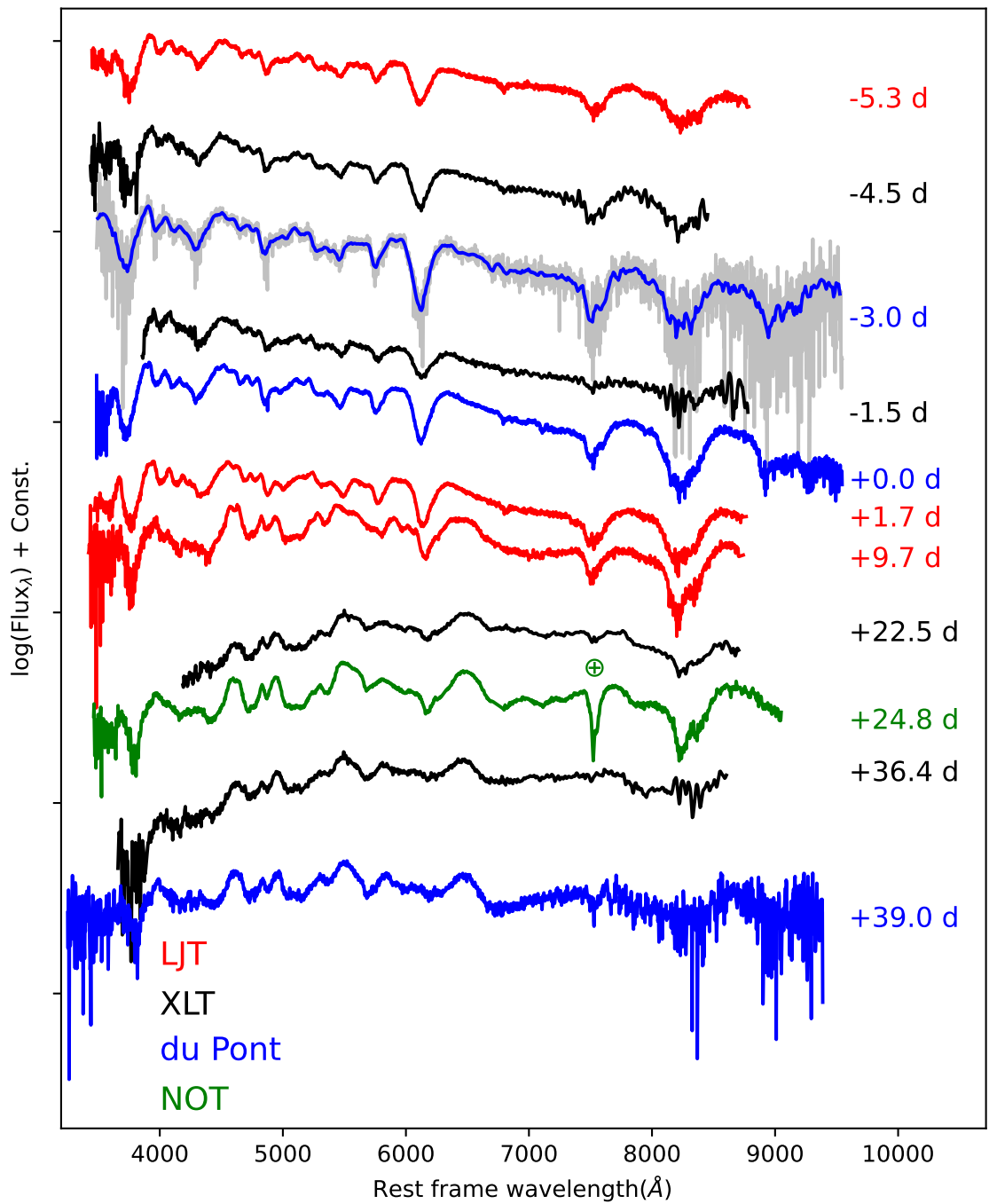}
\caption{Optical spectral evolution of SN 2012ij. The t $\sim$ $-$3day spectrum with worse signal-to-noise ratio is resampled with a bin of 15 Angstrom, and the original spectrum is plotted in gray. The spectra are shifted vertically for the clarity. The epochs relative to the $B$-band maximum light are labelled on the right. Strongest telluric region is marked with $\oplus$.}
\label{spectra_12ij}
\end{figure}

In Figure \ref{spectra_12ij}, a total of 11 optical spectra are shown for SN 2012ij, spanning the phases from t $\sim~-$5 days to t $\sim$ +39 days relative to the $B$-band maximum light. One can see that SN 2012ij shares some common evolutionary features with the sub-class of 91bg-like SNe Ia. The pre-maximum spectra showed prominent P-Cygni features of intermediate-mass elements (IME), including Si~{\sc ii}, S~{\sc ii}, Ca~{\sc ii} and Mg~{\sc ii}. The Ti~{\sc ii} absorption feature, which is characteristic of the 91bg-like subclass, is also prominent in SN 2012ij, confirming its classification as a subluminous SN Ia. Moreover, the Si~{\sc ii} $\lambda$5972 and the O~{\sc i} $\lambda$7774 absorption features were found to be much stronger than normal SNe Ia at similar epochs. After the maximum light, the continuum became flat quickly, which is consistent with the color-curve evolution. The S~{\sc ii} lines almost disappeared at about one week after $B$-band maximum, and the Si~{\sc ii} $\lambda$5972 absorption feature became undetectable at t $\sim$ +20 days. The Si~{\sc ii} $\lambda$6355 absorption line was still detectable in the t $\sim$ +25d spectrum, but it became invisible in the t $\sim$ +36d spectrum. On the contrary, some other spectral features like `Ti trough' and IGE lines around 5000~{\AA} tended to become stronger after the maximum light.

\begin{figure}[ht]
\centering
\includegraphics[angle=0,width=140mm]{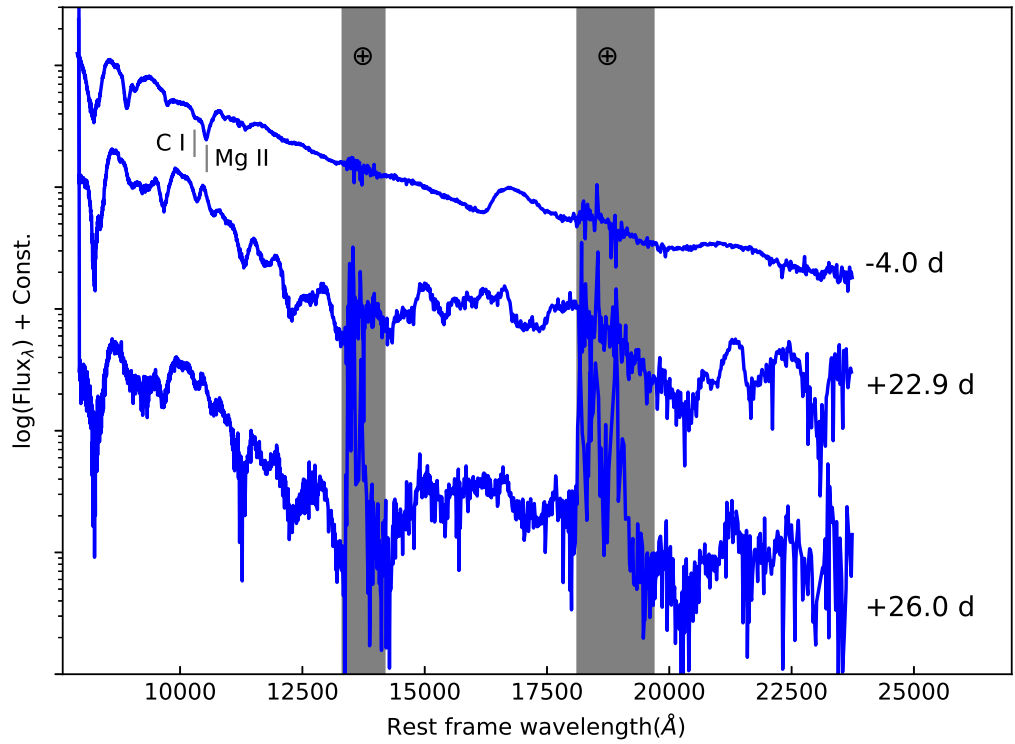}
\caption{NIR spectra of SN 2012ij from FIRE mounted on $Magellan$ $Baade$ Telescope. Strongest telluric region is marked with $\oplus$. All spectra are labelled with epoch with respect to the $B$-band maximum on the right. The spectra are shifted vertically for clarity.}
\label{NIR}
\end{figure}

The NIR spectra of SN 2012ij, obtained at t $\sim$ $-$4, +23, and +26 days relative to the B-band maximum, are displayed in Figure \ref{NIR}. At t $\sim$ $-$4d, the spectrum was characterized by some absorption lines of intermediate-mass elements (IMEs) like Mg~{\sc ii}, C~{\sc i}, O~{\sc i} and Si~{\sc ii}; the velocity inferred from C~{\sc i} $\lambda$1.0693 $\mu$m absorption is about 10900 $\pm$ 300 km s$^{-1}$, which is significantly lower than that of SN 1999by ($\sim$ 13000 km s$^{-1}$). The strength of C~{\sc i} absorption is weaker than that of SN 1999by \citep{hof02} but a bit stronger than those of the transitional objects iPTF13ebh \citep{hsi15} and SN 2015bp \citep{wya21}. All these supernovae with C~{\sc i} $\lambda$1.0693 $\mu$m absorption features are subluminous, which could be due to that neutral carbon only exists in low-luminosity SNe Ia. The existence of C~{\sc i} $\lambda$1.0693 $\mu$m helps constrain the explosion models, which will be discussed in section \ref{sectionModel}. At t $\sim$ 3 weeks after the peak, the IME lines almost disappeared in the spectra of SN 2012ij and the features were dominated by Ca~{\sc ii}, Fe~{\sc ii}, Co~{\sc ii} and Ni~{\sc ii} lines with the cooling of the photosphere, SN 1999by displayed similar spectral evolution at this similar phase \citep{hof02}. 
At t $\sim$ +23 days, prominent iron-peak emission in $H$-band were detected, of which the blue-edge velocity $v_{edge}$ can determine the outer location of Ni. Following the method described by \citet{ash19a}, we measured $v_{edge}$ of SN 2012ij as 5400 $\pm$ 400 km s$^{-1}$ at t $\sim$ +23d, which is comparable to that of the transitional object iPTF13ebh at similar phase.

\begin{figure}[ht]
\centering
\includegraphics[angle=0,width=160mm]{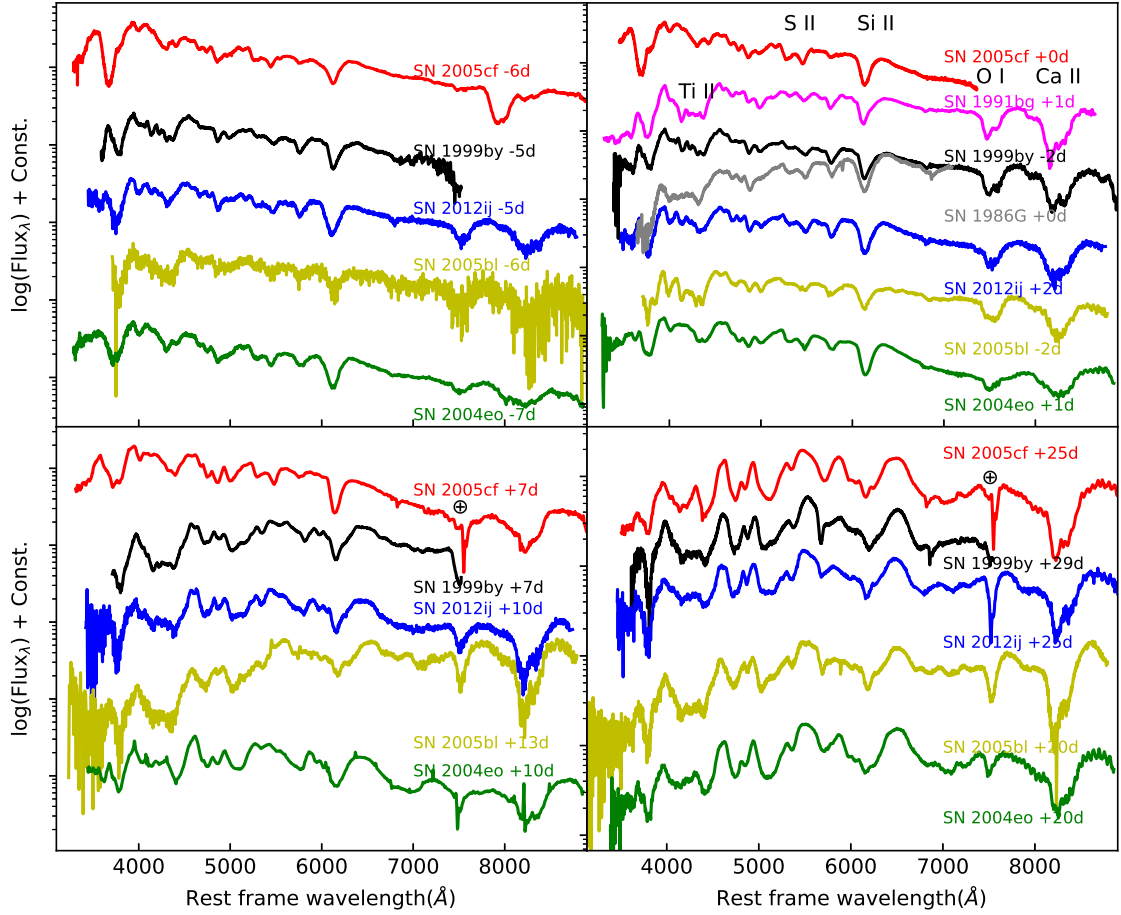}
\caption{Optical spectra of SN 2012ij at four different epochs (around t $\sim$ $-$5d, +2d, +10d and +25d) compared to other subluminous SNe 1991bg, 1999by and 2005bl, normal type Ia SN 2005cf, and transitional SN 2004eo. All spectra have been corrected for redshift and the epochs relative to $B$-band maximum light are labelled on the right. Strongest telluric region is marked with $\oplus$.}
\label{spec_compare1}
\end{figure}

Figure \ref{spec_compare1} shows detailed comparisons of the optical spectra of SN 2012ij with other sub-luminous and normal SNe Ia at several epochs (i.e. at t $\sim$ $-$5d, +2d, +10d and +25days). The comparison sample include SNe 1986G \citep{ham02}, 1991bg \citep{lei93,tur96}, 1999by \citep{gar04}, 2004eo \citep{pas07b}, 2005bl \citep{tau08} and 2005cf \citep{wxf09b}. Like SNe 1986G and 1999by, SN 2012ij did not exhibit a strong Ti absorption trough near the maximum light. After t $\sim$ +10d, SN 2012ij showed significant absorption features of Ti~{\sc ii} and other IGE. At each epoch in comparison, one can see that SN 2012ij showed close resemblance to SN 1999by in both overall shapes and strength of characteristic spectral lines.

\subsection{Optical Spectroscopic Properties}

The pEWs of absorption features near 5750 {\AA} and 6100 {\AA}, attributed to the Si~{\sc ii} $\lambda$5972 and the Si~{\sc ii} $\lambda$6355, respectively, have been used to classify SNe Ia \citep{bran06,bran09}. For SN 2012ij, the pEWs of these two spectral features were measured as 54 $\pm$ 3 {\AA} and 109 $\pm$ 4 {\AA}, respectively. These two values are quite similar to those derived from SN 1999by (i.e., pEW(Si~{\sc ii} $\lambda$5972) = 53 \AA, pEW(Si~{\sc ii} $\lambda$6355) = 107 \AA; \citealt{blond12}), suggesting that SN 2012ij belongs to the `cool' subclass. 

\begin{figure}[ht]
\centering
\includegraphics[angle=0,width=140mm]{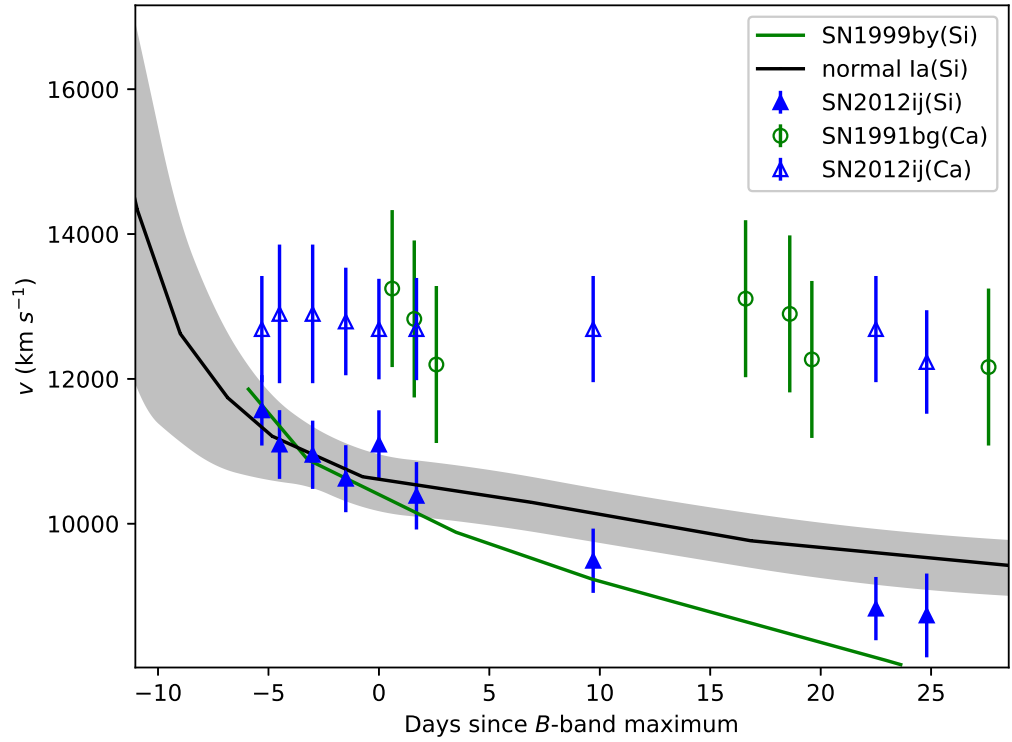}
\caption{Photospheric velocity evolution of SN 2012ij measured by the absorption minimum of Si~{\sc ii} $\lambda$6355, compared with 91bg-like and normal SNe Ia, with the mean Si~{\sc ii} evolution for normal SNe Ia (dark line) and 91bg-like events (green line) overplotted. The gray region represents 1$\sigma$ uncertainty for the Si~{\sc ii} velocity of normal SNe Ia \citep{wxf09a}. 
The velocities of Ca~{\sc ii} triplet measured for SN 1991bg and SN 2012ij are overplotted for comparison.}
\label{vSi}
\end{figure}

For SNe Ia, the Si~{\sc ii} $\lambda$6355 absorption feature is often used to infer the photospheric velocity, which was measured as $v_{\rm{Si}}$ = 10,500 $\pm$ 300 km s$^{-1}$ for SN 2012ij around the $B$-band maximum light. We show the velocity evolution of SN 2012ij and other SNe Ia in Figure \ref{vSi}, where one can see that SN 2012ij and SN 1999by share similar photospheric velocity near maximum light and velocity evolution until t $\sim$ 20 days.
According to the definition by \citet{ben05}, the velocity gradient measured for SN 2012ij is 87 $\pm$ 8 km s$^{-1}$ d$^{-1}$, which is close to other 91bg-like SNe (i.e., SN 1997cn; 83 $\pm$ 10 km s$^{-1}$ d$^{-1}$ \citealt{ben05}). 

Besides Si~{\sc ii} $\lambda$6355, we also used Ca~{\sc ii} triplet absorption feature near 8583 {\AA} to estimate the ejecta velocity $v_{\rm{Ca}}$. The velocity inferred from Ca~{\sc ii} triplet of SNe 2012ij is higher than Si~{\sc ii} and it varies slowly with time since t $\sim$ $-$5 days, which is similar to that seen in other 91bg-like SNe \citep{fol13}. No detached high-velocity features of Si~{\sc ii}, Ca~{\sc ii} or O~{\sc i}, as described in \citet{zhao15,zhao16}, were detected in SN 2012ij.

\section{Discussion}\label{discussion}

\subsection{Transitional Light Curve Properties}\label{discussion_lc}

SN 2012ij is a fast decliner with ${\Delta}m_{15}(B)$ = 1.86 mag and its $i$-band primary maximum occurred 0.06 days before the $B$-band maximum, which matches properties of transitional SNe Ia \citep{hsi15}. Unlike 91bg-like SNe, prominent secondary maximum features are found in the $I$- band light curves of SN 2012ij , which has long been proposed as one of the most important criteria of distinguishing intrinsically peculiar 91bg-like SNe Ia and fast-declining (or cooler) SNe Ia \citep{dha17}. Stronger secondary maximum feature in NIR bands usually appears in normal SNe Ia, which is proposed to be related to the recombination of IGE \citep{kas06}. \citet{tau17} suggested that this secondary maximum/shoulder feature usually appear earlier and weaker in cooler SNe, where the recombination of IGE would be earlier and the NIR double peaks could merge to form a delayed single peak as seen in some 91bg-like SNe. The transitional SNe Ia exhibit luminosity and temperature lying between normal and 91bg-like SNe Ia, and it is reasonable that their NIR-band light curves might form weak secondary maximum features.

We adopt three parameters to describe the properties of the NIR secondary peak. The first parameter is about the recombination time of IGE, which can be estimated by the time when the $B-V$ color curve shows a break $t_{BV}$ \citep{wyg19}. Here we use $s_{BV}$, a parameter similar to $t_{BV}$, to quantify the time of secondary peak. For subluminous SNe Ia, the primary peaks in $i$ band could be delayed due to merging with the secondary ones. Thus, the second parameter t$^{i-B}_{\rm{max}}$, the time of $i$-band maximum relative to that of the $B$-band maximum, can be used to evaluate the degree of merging. To better quantify the intensity of the $i$-band secondary maximum, we adopt the third parameter $\mathcal{\overline{F}}_{i2}$, which is calculated by dividing the flux integrated over interval from $t=+15$ to $t=+40$ days with the $i$-band flux at the primary peak. This parameter is similar to those proposed by \citet{kri01}, \citet{burns14} and \citet{pap19}, but we have chosen a time interval relative to the $i$-band maximum instead of $B$-band maximum due to the large variation of t$^{i-B}_{\rm{max}}$ between normal and subluminous SNe Ia.

\begin{figure}[ht]
\centering
\includegraphics[angle=0,width=180mm]{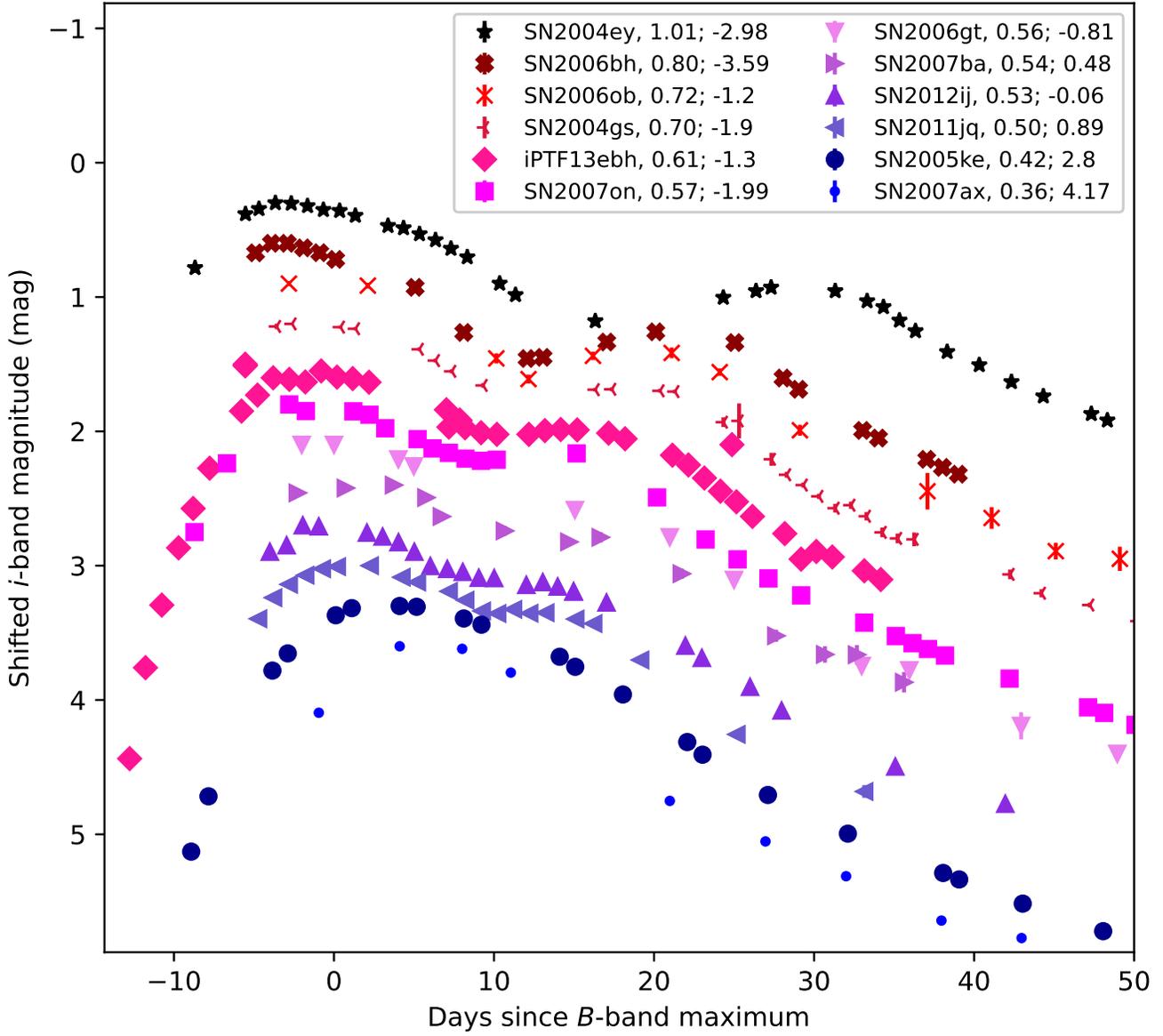}
\caption{$i$-band light curves marked in red to blue based on their $s_{BV}$. Typical 91bg-like SNe Ia (marked with circles) shows little secondary peak at the bottom of the figure, and normal SNe Ia (marked with crosses) shows strong secondary peaks at the top. Between them, the secondary peaks of transitional SNe (triangles and squares) are found to appear later and stronger with the rise of $s_{BV}$ and the decline of t$^{i-B}_{\rm{max}}$. Transitional SNe with similar light curves to SN 2012ij are marked with triangles. Peak magnitudes are normalized. Their $s_{BV}$ and t$^{i-B}_{\rm{max}}$ (days) are listed after their names.}
\label{pic:trans_lc}
\end{figure}

From the CSP sample analyzed by \citet{ash20}, we find that SNe 2006gt, 2007ba, 2011jq have similar $s_{BV}$ and t$^{i-B}_{\rm{max}}$ with SN 2012ij. To match the CSP sample, we only discuss the features of secondary maximum in $i$ band. The $i$-band light curves of three comparison SNe Ia \citep{con10,wal15} are shown in Figure \ref{pic:trans_lc}. The main parameters of these SNe Ia are shown in Table \ref{table:trans}. We noticed that these SNe having similar $s_{BV}$ and t$^{i-B}_{\rm{max}}$ as SN 2012ij also show similar light curve morphology.

\begin{figure}[ht]
\centering
\includegraphics[angle=0,width=160mm]{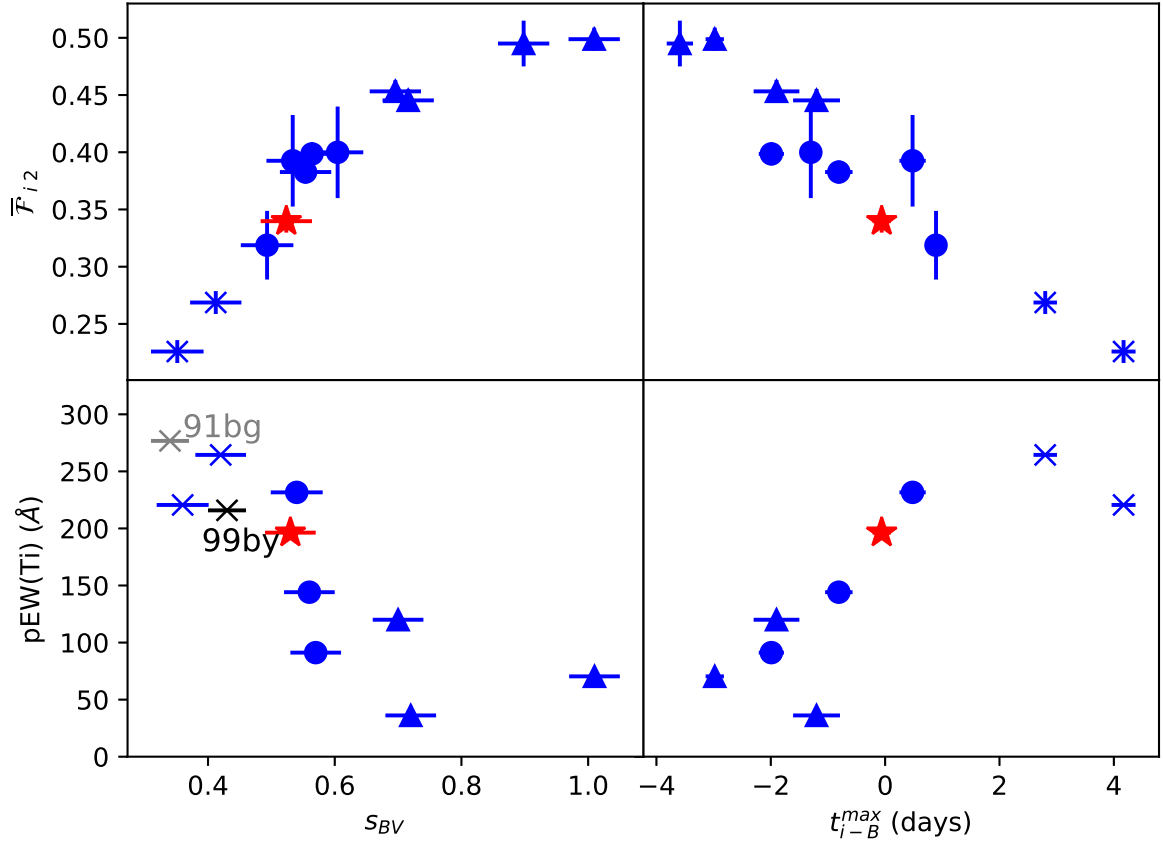}
\caption{Relation of the mean normalized flux of $i$-band secondary maximum ($\mathcal{\overline{F}}_{i2}$; upper panel), the pEW of Ti absorption lines near maximum light (pEW(Ti); lower panel), with $s_{BV}$ (left) and t$^{i-B}_{\rm{max}}$ (right) of samples in Table \ref{table:trans}. The data of SN 2012ij are identified with red star. Normal (triangle), 91bg-like (cross), and transitional (circle) SNe are marked with different symbols.}
\label{pic:sbv-tib}
\end{figure}

We plot $\mathcal{\overline{F}}_{i2}$ against $s_{BV}$ and t$^{i-B}_{\rm{max}}$ in the upper panel of Figure \ref{pic:sbv-tib}, where $\mathcal{\overline{F}}_{i2}$ shows clear correlations with both $s_{BV}$ and t$^{i-B}_{\rm{max}}$. These correlations imply that the intensity of the secondary maximum should have a continuous distribution and be sensitive to the occurrence time of the secondary maximum as well as the merging degree of the $i$-band double peaks. We also note that, the correlation of $\mathcal{\overline{F}}_{i2}$ with t$^{i-B}_{\rm{max}}$ deviates from the correlation between $\mathcal{\overline{F}}_{i2}$ and $s_{BV}$ for normal SNe Ia. This deviation may be attributed to the fact that for normal SNe Ia, the double peaks are detached so that t$^{i-B}_{\rm{max}}$ is decoupled from the double-peak merging. The same phenomenon was also presented between t$^{i-B}_{\rm{max}}$ and $s_{BV}$. When $s_{BV}<$ 0.8, t$^{i-B}_{\rm{max}}$ shows a negative correlation with $s_{BV}$ (see Figure 3 of \citealt{ash20}), whereas for normal SNe Ia (i.e., $s_{BV}>$ 0.8), the correlation is not obvious.

\subsection{Other Transitional Properties}\label{discussion_tran}

\begin{figure}[ht]
\centering
\includegraphics[angle=0,width=180mm]{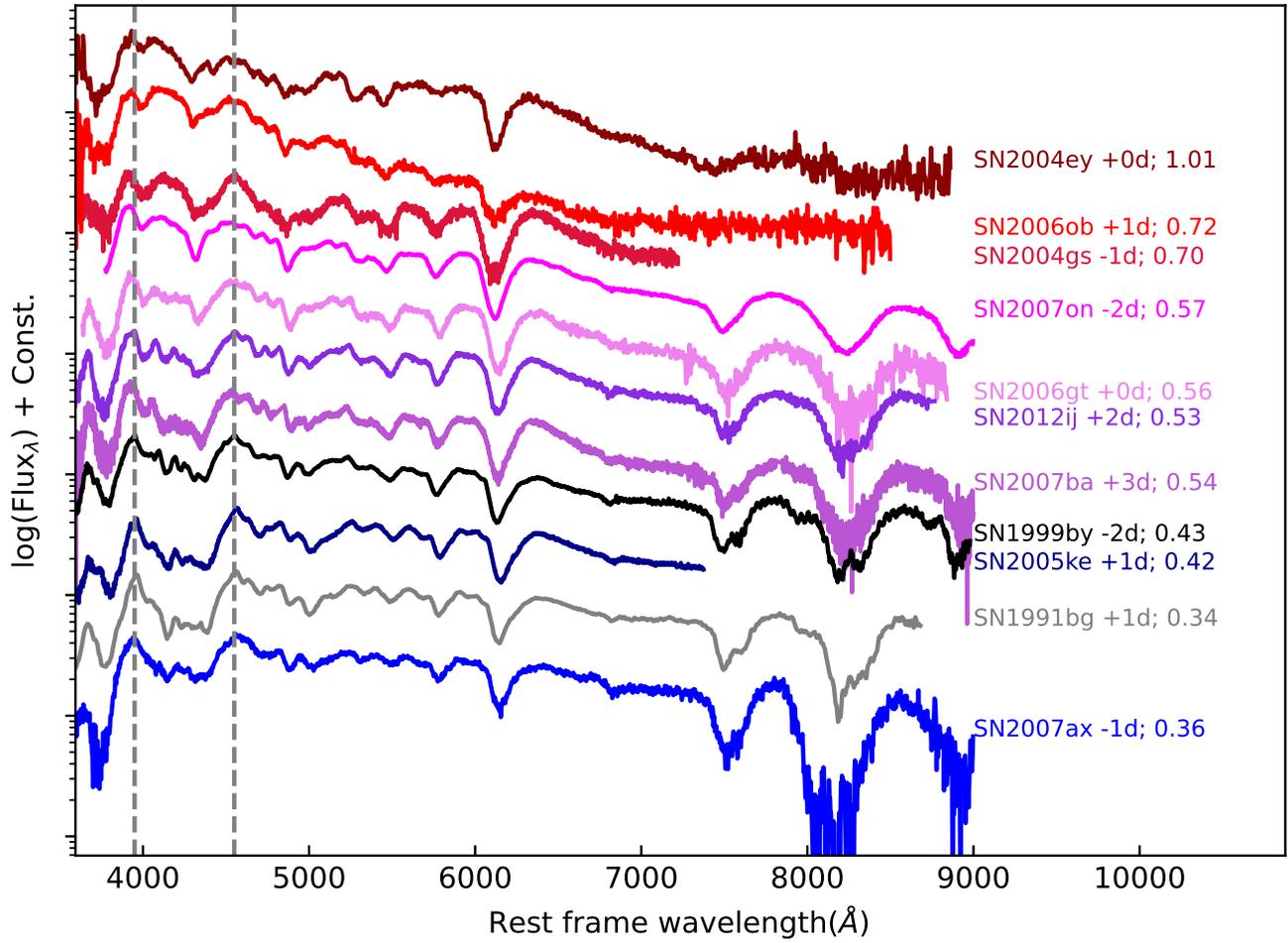}
\caption{Near-maximum light spectra of SNe Ia listed in Table \ref{table:trans}. These spectra displayed in order of decreasing $s_{BV}$ parameter.}
\label{pic:trans_spec}
\end{figure}

In Figure \ref{pic:trans_spec}, we compare the near-maximum-light spectra of a larger SN Ia sample as listed in Table \ref{table:trans}. We measured pEW of Ti~{\sc ii} absorption, corresponding to the equivalent width from 3950 to 4550\AA, to quantify the intensity of characteristic Ti absorption features. Though this feature is strongly contaminated by other absorption lines (i.e., Si~{\sc ii}, Mg~{\sc ii} and Fe~{\sc iii} line features), it can still be an effective parameter roughly reflecting the strength of Ti~{\sc ii} trough. We plot the pEW(Ti~{\sc ii}) of all these spectra in the lower panel of Figure \ref{pic:sbv-tib}, and find that the strength of Ti~{\sc ii} absorption trough of these sub-luminous SNe Ia is also correlated with their $s_{BV}$ and t$^{i-B}_{\rm{max}}$, suggesting that the Ti~{\sc ii} trough is also sensitive to the secondary maximum and merging degree of the NIR double peaks.

We also noticed that SNe 2006gt, 2007ba and 2011jq with comparable $s_{BV}$ (0.50 - 0.56) and t$^{i-B}_{\rm{max}}$ ($-$0.81 - 0.89 days) to SN 2012ij show similar light curves and spectra. Here we call the SNe with $s_{BV}$ ranging from 0.50 to 0.56 and t$^{i-B}_{\rm{max}}$ ranging from $-$0.81 to 0.89 days as 12ij-like SNe for convenience, but this in no way considers them to be a separate subclass. The percentage of 12ij-like SNe is about $2\%$ in the published sample of SNe Ia from the CSP-II \citep{ash20}.

A widely recognized transitional SN 1986G, of which $s_{BV}$ = 0.54 falls in the range of 0.50 - 0.56 but $i$-band light curve is not obtained, also shows similar light curves and spectra to SN 2012ij. It implies that SN 1986G is also a member of 12ij-like SNe, and classification of 12ij-like SNe should partially overlap with previous classifications of transitional SNe Ia.

Besides, since the percentage and sample size of 12ij-like/transitional SNe are comparable to the 91bg-like subclass (see Figure 3 and Table 2 in \citealt{ash20}), a continuous distribution is more reasonable. Moreover, in other statistical studies, continuous distribution in different parameter spaces have also been found between normal and 91bg-like SNe Ia \citep{gon11,blond12,burns14,burns18}.

Note that all these four transitional SNe Ia in our sample are located in passive galaxies. The host galaxies of SNe 2007ba, 2011jq, 2012ij are S0/a galaxies, while SN 2006gt exploded in an elliptical galaxy. \citet{liw11a} showed that only 91bg-like SNe Ia strongly prefer to occur in early-type galaxies, which implies these transitional SNe Ia are similar to 91bg-like SNe Ia instead of normal SNe Ia in terms of host-galaxy environments.

\subsection{Model Comparison}\label{sectionModel}

To constrain the possible explosion mechanism of these transitional objects, we compare the observed spectra of SN 2012ij with synthetic spectra predicted from two promising models: the $M_{\rm{Ch}}$ delayed-detonation model and the sub-$M_{\rm{Ch}}$ model. The $M_{\rm{Ch}}$ delayed-detonation model (hereafter called the DDT model; \citealt{hoe17}) gives good predictions of both light curves and spectra for normal SNe Ia, and this model has also been proposed to be able to produce light/color curves of 91bg-like SNe Ia. The sub-$M_{\rm{Ch}}$ explosions, usually realized as violent mergers and double detonations, gives narrow, fast-evolving light curves, which can fit the observations of 91bg-like SNe Ia as well \citep{pak13}. Here we adopt the pure central detonation of sub-$M_{\rm{Ch}}$ WD (hereafter called the SCH model; \citealt{blond17}), which has similar properties to the sub-$M_{\rm{Ch}}$ model with a thin He-shell. Both the DDT and the SCH models can predict explosions of SNe Ia with light curve properties ranging from normal to 91bg-like SNe Ia \citep{fin10,hoe17,blond17}.

\begin{figure}[ht]
\centering
\includegraphics[angle=0,width=180mm]{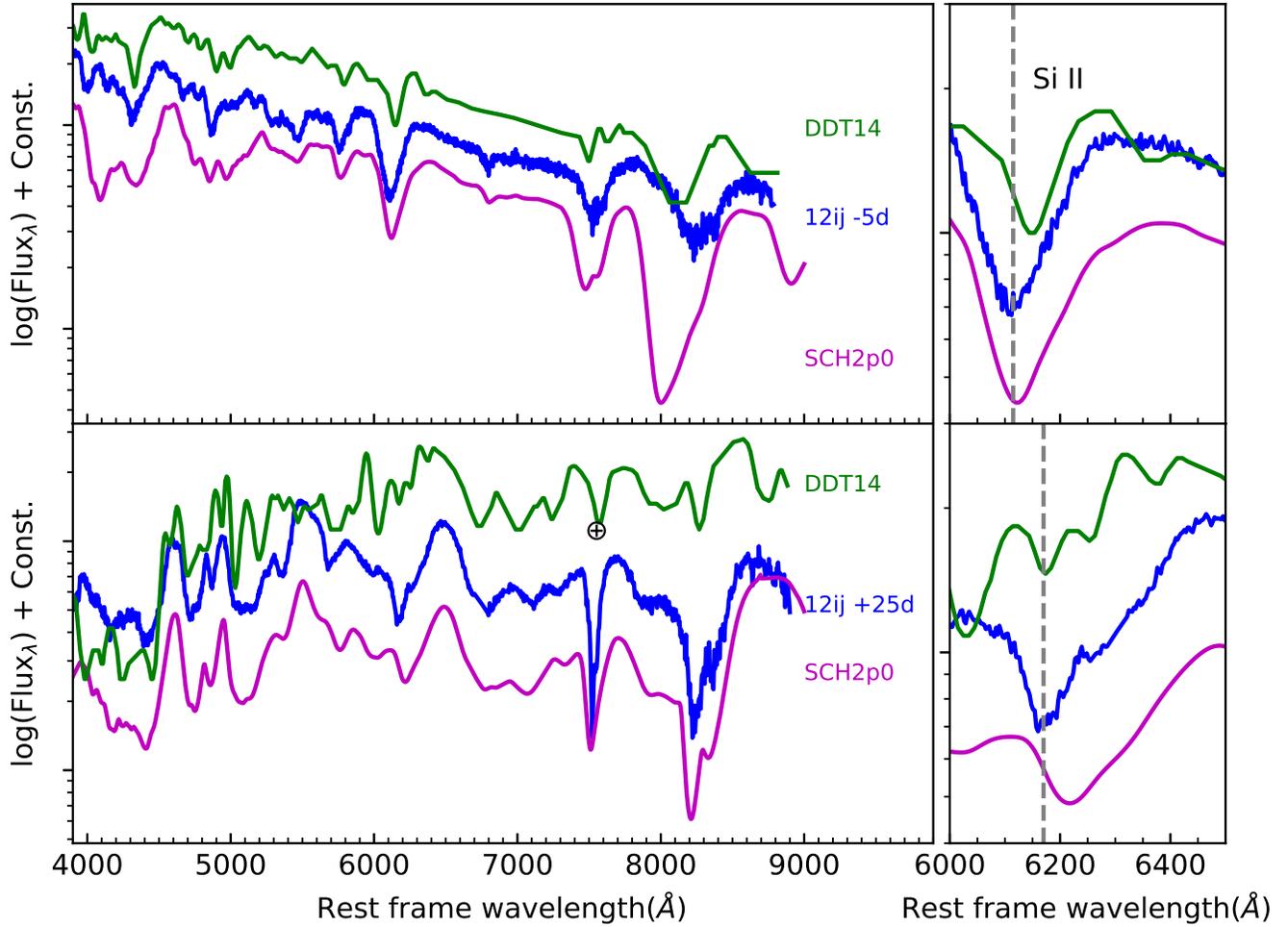}
\caption{\textit{Left}: Optical spectra of SN 2012ij compared with two synthetic spectra from DDT14 and model SCH2p0 (see Table \ref{table:models}) at two different epochs: $-$5 days (top panel) and +25 days (bottom panel) relative to $B$-band maximum light. All spectra have been corrected for redshift. \textit{Right}: the zoomed areas of Si~{\sc ii} $\lambda$6355 absorption features. The gray line corresponds to the Si~{\sc ii} absorption minimum of SN 2012ij.}
\label{spec_compare3}
\end{figure}

\begin{figure}[ht]
\centering
\includegraphics[angle=0,width=180mm]{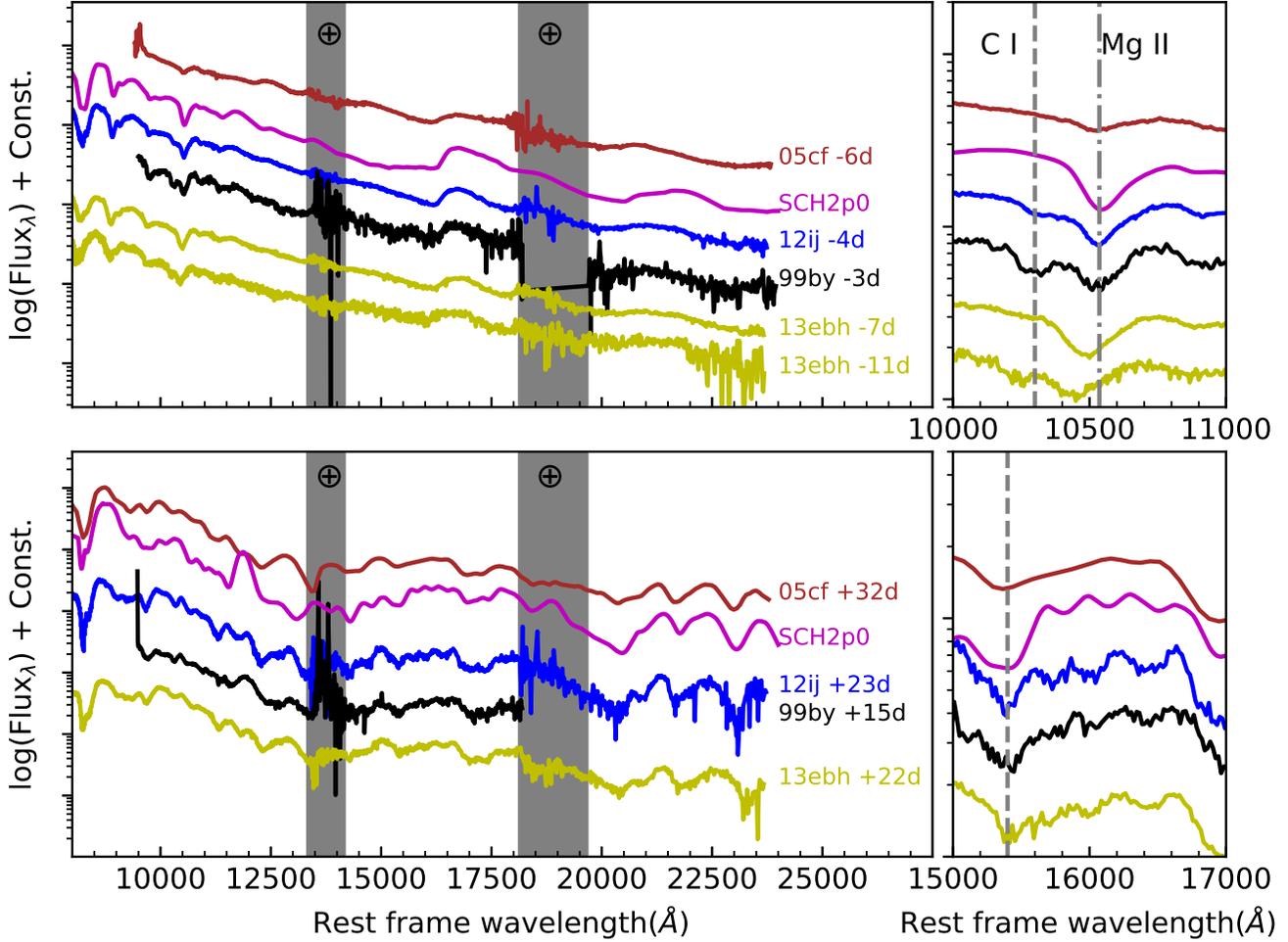}
\caption{\textit{Left}: Near-infrared spectra of SN 2012ij compared to those of two sub-luminous SNe iPTF13ebh and 1999by, normal type Ia SN 2005cf and synthetic spectrum of models SCH2p0 (see Table \ref{table:models}) obtained at t $\sim$ a week before maximum light (top panel) and t $\sim$ 20 days after maximum light (bottom panel). All spectra have been corrected for host-galaxy redshift and the epochs relative to the $B$-band maximum light are labelled on the right. \textit{Top right}: the zoomed area of C~{\sc i} $\lambda$1.0693 $\mu$m and Mg~{\sc ii} absorption feature of the spectra near maximum light. The gray dashed and dash-dotted lines represent the positions of corresponding absorptions in SN 2012ij. \textit{Bottom right}: the zoomed area of $H$-band region t $\sim$ 20 days after maximum light. The gray dashed line represent the blue-edge of the +23d spectra for SN 2012ij (5,400 km s$^{-1}$).}
\label{spec_compare2}
\end{figure}

In Figures \ref{spec_compare3} and \ref{spec_compare2}, we compare the optical and NIR spectra of SN 2012ij with the synthetic spectra yielded from model 14 of \citet{hoe17} (hereafter DDT14) and model SCH2p0 of \citet{blond17}. These two models are chosen due to that their light curve parameters are similar to those of SN 2012ij. The main parameters of these models are listed in Table \ref{table:models}.

As shown in Figure \ref{spec_compare3}, the synthetic spectrum from DDT14 model shows better conformation to that of SN 2012ij taken at t = $-$5 days, especially the Ti~{\sc ii} trough and the Ca NIR triplet. In comparison, the SCH2p0 model could better reproduce the Si~{\sc ii} profile. Moreover, the latter model seems to fit well the spectrum of SN 2012ij taken at t = +25 days. Figure \ref{spec_compare2} shows the comparison of NIR spectra of SN 2012ij with SNe 1999by, 2005cf, iPTF13ebh, and the model spectrum of SCH2p0. One can see that SN 2012ij is overall similar to SN 1999by. At t = $-$4 days, the NIR spectrum of SN 2012ij shows weak C~{\sc i} $\lambda$1.0693 $\mu$m absorption, which is similar to those seen in SN 1999by and iPTF13ebh. The DDT model can produce this carbon line \citep{hof02, hsi15}, but it is not detected in the spectrum yielded from model SCH2p0 at similar phases. However, the SCH model does not predict a lot amount of carbon in the outer ejecta because of surface detonation \citep{pol19}. However, \citet{liw21} suggested that a small amount of residual carbon could be due to an orientation-dependent effect, resulted from an asymmetric explosion in helium detonation scenario. At t $\sim$ 3 weeks after the maximum light, the $v_{edge}$ in t $\sim$ +23d spectrum of SN 2012ij is similar to that of SN 1999by, but slightly higher than that of SCH2p0. \citet{ash19b} showed that the NIR spectra of SN 1999by match well with those produced by the model DDT08 \citep{hof02}.

In comparison, the DDT model can predict reasonable NIR C~{\sc i} $\lambda$1.0693 $\mu$m feature and velocity of Ni edge in $H$ band, but the Si~{\sc ii} velocity is underestimated by this model, which might be eliminated by changing the central density \citep{hoe17}. 
On the other hand, the SCH model can also generate the optical spectra similar to those of SN 2012ij. However, whether this model can produce proper C~{\sc i} feature in the NIR spectra still remains controversial \citep{pol19,liw21}. 

In conclusion,the main portion of the spectral features ranging from normal to 91bg-like SNe Ia can be reproduced by the DDT and the SCH models. Therefore, some of the diversities observed between 91bg-like and normal SNe Ia may be attributed to different initial parameters rather than to different explosion models.

\section{Conclusion}\label{conclusion}

In this paper, we present and analyze the photometric and spectroscopic data of a peculiar, subluminous SN Ia 2012ij. This SN is found to be very similar to SN 1999by in many aspects, including the color-stretch factor $s_{BV}$, bolometric light curves, optical and NIR spectra, spectral parameter (e.g., pEW(Si~{\sc ii} $\lambda$6355\&5972), and velocities inferred from Si~{\sc ii} $\lambda$6355 and Ca NIR triplet). 

By comparing parameters of light curves and spectra, we found that $\sim$2\% SNe in CSP-II SNe Ia sample can be identified as the subclass like SN 2012ij, which exhibit transitional light-curve features linking normal to 91bg-like SNe Ia. Through some quantitative analysis, we found that the time of NIR primary peak t$^{i-B}_{\rm{max}}$, the relative intensity of NIR secondary peak $\mathcal{\overline{F}}_{i2}$, the time when IGEs recombine, and the color-stretch factor $s_{BV}$ are all correlated with the extent of merging between two peaks in NIR bands, which may explain for the light-curve diversity of SNe Ia in NIR bands. 

We show both the $M_{\rm{Ch}}$ delayed-detonation model and the sub-$M_{\rm{Ch}}$ model have the potential to re-produce the observations of SN 2012ij. Despite the DDT models could produce the entire observed range from normal to 91bg-like SNe Ia \citep{hoe17}, the SCH model also seems to work for most SNe Ia except for the fastest declining objects like SN 1991bg \citep{blond17}. This reproducibility of a single model for various SNe Ia observations, together with clear correlations and continuous distributions in spectral and light curve parameters, suggest that the existence of a continuous transition from normal to 91bg-like SNe Ia.

With tons of multi-color light curves from the upcoming Rubin Observatory Legacy Survey Telescope (LSST; \citealt{LSST}), the subclass like SN 2012ij will be greatly enriched. This subclass could be recognized by similar photometric parameters to SN 2012ij in this work, allowing further identification of their association with normal and 91bg-like SNe Ia. Moreover, statistical studies of vast numbers of early light/color curves may be able to identify diversities that cannot be detected at a later stage. This will help us find more relations between their properties and hence better understand their progenitors and explosion mechanism.

\acknowledgments

We acknowledge the support of the staff of the Xinglong 2.16m and 80cm telescope. This work was partially supported by the Open Project Program of the Key Laboratory of Optical Astronomy, National Astronomical Observatories, Chinese Academy of Sciences.

This work is supported by the National Natural Science Foundation of China (NSFC; grant Nos. 11890691, 11890693, 11733007, 11673027, 11873053, 12073035, 12033003, and 11633002). This work is also supported by the National Program on Key Research and Development Project (grant no. 2016YFA0400803) and China Manned Spaced Project (CMS-CSST-2021-A12). This work is partially supported by the Scholar Program of Beijing Academy of Science and Technology (DZ: BS202002). J.-J. Zhang is supported by the NSFC (grants 11403096, 11773067), the Key Research Program of the CAS (Grant NO. KJZD-EW-M06), the Youth Innovation Promotion Association of the CAS (grants 2018081), and the CAS “Light of West China” Program. Funding for the LJT has been provided by Chinese Academy of Sciences and the People’s Government of Yunnan Province. The LJT is jointly operated and administrated by Yunnan Observatories and Center for Astronomical Mega-Science, CAS. BW is supported by the Yunnan Fundamental Research Projects (Nos 2018FB005X, 2019FJ001 and 202001AS070029). The work of the CSP-II has been generously supported by the National Science Foundation under grants AST-1008343, AST-1613426, AST-1613455, and AST-1613472. M. S. is funding in part by a research Project 1 grant from the Independent Research Fund Denmark (IRFD grant number 13261) and by a VILLUM FONDEN Experiment (grant number 28021).

We thank P. A. Mazzali, C. Wu, D. Liu, W. Li, Q. Jiang, Y. Zhang for helpful discussion on this paper.

\software{SWARP\ \ \citep{SWARP},\ \ SExtractor\ \ \citep{SExtractor},\ \ SCAMP\ \ \citep{SCAMP},\ \ Matplotlib\ \ \citep{matplotlib},\ \ NumPy\ \ \citep{numpy},\ \ SciPy\ \ \citep{scipy}.}

\clearpage

\bibliographystyle{aasjournal}
\bibliography{lizt}

\begin{thebibliography}{}
\expandafter\ifx\csname natexlab\endcsname\relax\def\natexlab#1{#1}\fi
\providecommand{\url}[1]{\href{#1}{#1}}
\providecommand{\dodoi}[1]{doi:~\href{http://doi.org/#1}{\nolinkurl{#1}}}
\providecommand{\doeprint}[1]{\href{http://ascl.net/#1}{\nolinkurl{http://ascl.net/#1}}}
\providecommand{\doarXiv}[1]{\href{https://arxiv.org/abs/#1}{\nolinkurl{https://arxiv.org/abs/#1}}}

\bibitem[{{Arnett}(1982)}]{arn82}
{Arnett}, W.~D. 1982, \apj, 253, 785, \dodoi{10.1086/159681}

\bibitem[{{Ashall} {et~al.}(2019{\natexlab{a}}){Ashall}, {Hsiao}, {Hoeflich},
  {Stritzinger}, {Phillips}, {Morrell}, {Davis}, {Baron}, {Piro}, {Burns},
  {Contreras}, {Galbany}, {Holmbo}, {Kirshner}, {Krisciunas}, {Marion}, {Sand},
  {Shahbandeh}, {Suntzeff}, \& {Taddia}}]{ash19a}
{Ashall}, C., {Hsiao}, E.~Y., {Hoeflich}, P., {et~al.} 2019{\natexlab{a}},
  \apjl, 875, L14, \dodoi{10.3847/2041-8213/ab1654}

\bibitem[{{Ashall} {et~al.}(2019{\natexlab{b}}){Ashall}, {Hoeflich}, {Hsiao},
  {Phillips}, {Stritzinger}, {Baron}, {Piro}, {Burns}, {Contreras}, {Davis},
  {Galbany}, {Holmbo}, {Kirshner}, {Krisciunas}, {Marion}, {Morrell}, {Sand},
  {Shahbandeh}, {Suntzeff}, \& {Taddia}}]{ash19b}
{Ashall}, C., {Hoeflich}, P., {Hsiao}, E.~Y., {et~al.} 2019{\natexlab{b}},
  \apj, 878, 86, \dodoi{10.3847/1538-4357/ab204b}

\bibitem[{{Ashall} {et~al.}(2020){Ashall}, {Lu}, {Burns}, {Hsiao},
  {Stritzinger}, {Suntzeff}, {Phillips}, {Baron}, {Contreras}, {Davis},
  {Galbany}, {Hoeflich}, {Holmbo}, {Morrell}, {Karamehmetoglu}, {Krisciunas},
  {Kumar}, {Shahband eh}, \& {Uddin}}]{ash20}
{Ashall}, C., {Lu}, J., {Burns}, C., {et~al.} 2020, \apjl, 895, L3,
  \dodoi{10.3847/2041-8213/ab8e37}

\bibitem[{{Benetti} {et~al.}(2005){Benetti}, {Cappellaro}, {Mazzali},
  {Turatto}, {Altavilla}, {Bufano}, {Elias-Rosa}, {Kotak}, {Pignata}, {Salvo},
  \& {Stanishev}}]{ben05}
{Benetti}, S., {Cappellaro}, E., {Mazzali}, P.~A., {et~al.} 2005, \apj, 623,
  1011, \dodoi{10.1086/428608}

\bibitem[{{Bertin}(2006)}]{SCAMP}
{Bertin}, E. 2006, in Astronomical Society of the Pacific Conference Series,
  Vol. 351, Astronomical Data Analysis Software and Systems XV, ed.
  C.~{Gabriel}, C.~{Arviset}, D.~{Ponz}, \& S.~{Enrique}, 112

\bibitem[{{Bertin} \& {Arnouts}(1996)}]{SExtractor}
{Bertin}, E., \& {Arnouts}, S. 1996, \aaps, 117, 393,
  \dodoi{10.1051/aas:1996164}

\bibitem[{{Bertin} {et~al.}(2002){Bertin}, {Mellier}, {Radovich}, {Missonnier},
  {Didelon}, \& {Morin}}]{SWARP}
{Bertin}, E., {Mellier}, Y., {Radovich}, M., {et~al.} 2002, in Astronomical
  Society of the Pacific Conference Series, Vol. 281, Astronomical Data
  Analysis Software and Systems XI, ed. D.~A. {Bohlender}, D.~{Durand}, \&
  T.~H. {Handley}, 228

\bibitem[{{Betoule} {et~al.}(2014){Betoule}, {Kessler}, {Guy}, {Mosher},
  {Hardin}, {Biswas}, {Astier}, {El-Hage}, {Konig}, {Kuhlmann}, {Marriner},
  {Pain}, {Regnault}, {Balland}, {Bassett}, {Brown}, {Campbell}, {Carlberg},
  {Cellier-Holzem}, {Cinabro}, {Conley}, {D'Andrea}, {DePoy}, {Doi}, {Ellis},
  {Fabbro}, {Filippenko}, {Foley}, {Frieman}, {Fouchez}, {Galbany}, {Goobar},
  {Gupta}, {Hill}, {Hlozek}, {Hogan}, {Hook}, {Howell}, {Jha}, {Le Guillou},
  {Leloudas}, {Lidman}, {Marshall}, {M{\"o}ller}, {Mour{\~a}o}, {Neveu},
  {Nichol}, {Olmstead}, {Palanque-Delabrouille}, {Perlmutter}, {Prieto},
  {Pritchet}, {Richmond}, {Riess}, {Ruhlmann-Kleider}, {Sako}, {Schahmaneche},
  {Schneider}, {Smith}, {Sollerman}, {Sullivan}, {Walton}, \&
  {Wheeler}}]{bet14}
{Betoule}, M., {Kessler}, R., {Guy}, J., {et~al.} 2014, \aap, 568, A22,
  \dodoi{10.1051/0004-6361/201423413}

\bibitem[{{Blondin} {et~al.}(2018){Blondin}, {Dessart}, \& {Hillier}}]{blond18}
{Blondin}, S., {Dessart}, L., \& {Hillier}, D.~J. 2018, \mnras, 474, 3931,
  \dodoi{10.1093/mnras/stx3058}

\bibitem[{{Blondin} {et~al.}(2017){Blondin}, {Dessart}, {Hillier}, \&
  {Khokhlov}}]{blond17}
{Blondin}, S., {Dessart}, L., {Hillier}, D.~J., \& {Khokhlov}, A.~M. 2017,
  \mnras, 470, 157, \dodoi{10.1093/mnras/stw2492}

\bibitem[{{Blondin} {et~al.}(2012){Blondin}, {Matheson}, {Kirshner}, {Mand el},
  {Berlind}, {Calkins}, {Challis}, {Garnavich}, {Jha}, {Modjaz}, {Riess}, \&
  {Schmidt}}]{blond12}
{Blondin}, S., {Matheson}, T., {Kirshner}, R.~P., {et~al.} 2012, \aj, 143, 126,
  \dodoi{10.1088/0004-6256/143/5/126}

\bibitem[{{Branch} {et~al.}(2009){Branch}, {Chau Dang}, \& {Baron}}]{bran09}
{Branch}, D., {Chau Dang}, L., \& {Baron}, E. 2009, \pasp, 121, 238,
  \dodoi{10.1086/597788}

\bibitem[{{Branch} {et~al.}(1993){Branch}, {Fisher}, \& {Nugent}}]{bran93}
{Branch}, D., {Fisher}, A., \& {Nugent}, P. 1993, \aj, 106, 2383,
  \dodoi{10.1086/116810}

\bibitem[{{Branch} {et~al.}(2006){Branch}, {Dang}, {Hall}, {Ketchum},
  {Melakayil}, {Parrent}, {Troxel}, {Casebeer}, {Jeffery}, \& {Baron}}]{bran06}
{Branch}, D., {Dang}, L.~C., {Hall}, N., {et~al.} 2006, \pasp, 118, 560,
  \dodoi{10.1086/502778}

\bibitem[{{Brown} {et~al.}(2014){Brown}, {Breeveld}, {Holland}, {Kuin}, \&
  {Pritchard}}]{brown14}
{Brown}, P.~J., {Breeveld}, A.~A., {Holland}, S., {Kuin}, P., \& {Pritchard},
  T. 2014, \apss, 354, 89, \dodoi{10.1007/s10509-014-2059-8}

\bibitem[{{Burns} {et~al.}(2014){Burns}, {Stritzinger}, {Phillips}, {Hsiao},
  {Contreras}, {Persson}, {Folatelli}, {Boldt}, {Campillay}, {Castell{\'o}n},
  {Freedman}, {Madore}, {Morrell}, {Salgado}, \& {Suntzeff}}]{burns14}
{Burns}, C.~R., {Stritzinger}, M., {Phillips}, M.~M., {et~al.} 2014, \apj, 789,
  32, \dodoi{10.1088/0004-637X/789/1/32}

\bibitem[{{Burns} {et~al.}(2018){Burns}, {Parent}, {Phillips}, {Stritzinger},
  {Krisciunas}, {Suntzeff}, {Hsiao}, {Contreras}, {Anais}, {Boldt}, {Busta},
  {Campillay}, {Castell{\'o}n}, {Folatelli}, {Freedman}, {Gonz{\'a}lez},
  {Hamuy}, {Heoflich}, {Krzeminski}, {Madore}, {Morrell}, {Persson}, {Roth},
  {Salgado}, {Ser{\'o}n}, \& {Torres}}]{burns18}
{Burns}, C.~R., {Parent}, E., {Phillips}, M.~M., {et~al.} 2018, \apj, 869, 56,
  \dodoi{10.3847/1538-4357/aae51c}

\bibitem[{{Burrow} {et~al.}(2020){Burrow}, {Baron}, {Ashall}, {Burns},
  {Morrell}, {Stritzinger}, {Brown}, {Folatelli}, {Freedman}, {Galbany},
  {Hoeflich}, {Hsiao}, {Krisciunas}, {Phillips}, {Piro}, {Suntzeff}, \&
  {Uddin}}]{bur20}
{Burrow}, A., {Baron}, E., {Ashall}, C., {et~al.} 2020, \apj, 901, 154,
  \dodoi{10.3847/1538-4357/abafa2}

\bibitem[{{Cao} {et~al.}(2015){Cao}, {Kulkarni}, {Howell}, {Gal-Yam},
  {Kasliwal}, {Valenti}, {Johansson}, {Amanullah}, {Goobar}, {Sollerman},
  {Taddia}, {Horesh}, {Sagiv}, {Cenko}, {Nugent}, {Arcavi}, {Surace},
  {Wo{\'z}niak}, {Moody}, {Rebbapragada}, {Bue}, \& {Gehrels}}]{cao15}
{Cao}, Y., {Kulkarni}, S.~R., {Howell}, D.~A., {et~al.} 2015, \nat, 521, 328,
  \dodoi{10.1038/nature14440}

\bibitem[{{Cardelli} {et~al.}(1989){Cardelli}, {Clayton}, \& {Mathis}}]{car89}
{Cardelli}, J.~A., {Clayton}, G.~C., \& {Mathis}, J.~S. 1989, \apj, 345, 245,
  \dodoi{10.1086/167900}

\bibitem[{{Chonis} \& {Gaskell}(2008)}]{cho08}
{Chonis}, T.~S., \& {Gaskell}, C.~M. 2008, \aj, 135, 264,
  \dodoi{10.1088/0004-6256/135/1/264}

\bibitem[{{Contreras} {et~al.}(2010){Contreras}, {Hamuy}, {Phillips},
  {Folatelli}, {Suntzeff}, {Persson}, {Stritzinger}, {Boldt}, {Gonz{\'a}lez},
  {Krzeminski}, {Morrell}, {Roth}, {Salgado}, {Maureira}, {Burns}, {Freedman},
  {Madore}, {Murphy}, {Wyatt}, {Li}, \& {Filippenko}}]{con10}
{Contreras}, C., {Hamuy}, M., {Phillips}, M.~M., {et~al.} 2010, \aj, 139, 519,
  \dodoi{10.1088/0004-6256/139/2/519}

\bibitem[{{Dhawan} {et~al.}(2017){Dhawan}, {Leibundgut}, {Spyromilio}, \&
  {Blondin}}]{dha17}
{Dhawan}, S., {Leibundgut}, B., {Spyromilio}, J., \& {Blondin}, S. 2017, \aap,
  602, A118, \dodoi{10.1051/0004-6361/201629793}

\bibitem[{{Dilday} {et~al.}(2012){Dilday}, {Howell}, {Cenko}, {Silverman},
  {Nugent}, {Sullivan}, {Ben-Ami}, {Bildsten}, {Bolte}, {Endl}, {Filippenko},
  {Gnat}, {Horesh}, {Hsiao}, {Kasliwal}, {Kirkman}, {Maguire}, {Marcy},
  {Moore}, {Pan}, {Parrent}, {Podsiadlowski}, {Quimby}, {Sternberg}, {Suzuki},
  {Tytler}, {Xu}, {Bloom}, {Gal-Yam}, {Hook}, {Kulkarni}, {Law}, {Ofek},
  {Polishook}, \& {Poznanski}}]{dil12}
{Dilday}, B., {Howell}, D.~A., {Cenko}, S.~B., {et~al.} 2012, Science, 337,
  942, \dodoi{10.1126/science.1219164}

\bibitem[{{Dimitriadis} {et~al.}(2019){Dimitriadis}, {Foley}, {Rest}, {Kasen},
  {Piro}, {Polin}, {Jones}, {Villar}, {Narayan}, {Coulter}, {Kilpatrick},
  {Pan}, {Rojas-Bravo}, {Fox}, {Jha}, {Nugent}, {Riess}, {Scolnic}, {Drout},
  {K2 Mission Team}, {Barentsen}, {Dotson}, {Gully-Santiago}, {Hedges}, {Cody},
  {Barclay}, {Howell}, {KEGS}, {Garnavich}, {Tucker}, {Shaya}, {Mushotzky},
  {Olling}, {Margheim}, {Zenteno}, {Kepler spacecraft Team}, {Coughlin}, {Van
  Cleve}, {Cardoso}, {Larson}, {McCalmont-Everton}, {Peterson}, {Ross},
  {Reedy}, {Osborne}, {McGinn}, {Kohnert}, {Migliorini}, {Wheaton}, {Spencer},
  {Labonde}, {Castillo}, {Beerman}, {Steward}, {Hanley}, {Larsen},
  {Gangopadhyay}, {Kloetzel}, {Weschler}, {Nystrom}, {Moffatt}, {Redick},
  {Griest}, {Packard}, {Muszynski}, {Kampmeier}, {Bjella}, {Flynn},
  {Elsaesser}, {Pan-STARRS}, {Chambers}, {Flewelling}, {Huber}, {Magnier},
  {Waters}, {Schultz}, {Bulger}, {Lowe}, {Willman}, {Smartt}, {Smith}, {DECam},
  {Points}, {Strampelli}, {ASAS-SN}, {Brimacombe}, {Chen}, {Mu{\~n}oz},
  {Mutel}, {Shields}, {Vallely}, {Villanueva}, {PTSS/TNTS}, {Li}, {Wang},
  {Zhang}, {Lin}, {Mo}, {Zhao}, {Sai}, {Zhang}, {Zhang}, {Zhang}, {Wang},
  {Zhang}, {Baron}, {DerKacy}, {Li}, {Chen}, {Xiang}, {Rui}, {Wang}, {Huang},
  {Li}, {Cumbres Observatory}, {Hosseinzadeh}, {Howell}, {Arcavi}, {Hiramatsu},
  {Burke}, {Valenti}, {ATLAS}, {Tonry}, {Denneau}, {Heinze}, {Weiland},
  {Stalder}, {Konkoly}, {Vink{\'o}}, {S{\'a}rneczky}, {P{\'a}l}, {B{\'o}di},
  {Bogn{\'a}r}, {Cs{\'a}k}, {Cseh}, {Cs{\"o}rnyei}, {Hanyecz}, {Ign{\'a}cz},
  {Kalup}, {K{\"o}nyves-T{\'o}th}, {Kriskovics}, {Ordasi}, {Rajmon},
  {S{\'o}dor}, {Szab{\'o}}, {Szak{\'a}ts}, {Zsidi}, {ePESSTO}, {Williams},
  {Nordin}, {Cartier}, {Frohmaier}, {Galbany}, {Guti{\'e}rrez}, {Hook},
  {Inserra}, {Smith}, {Arizona}, {Sand}, {Andrews}, {Smith}, \&
  {Bilinski}}]{dimi19}
{Dimitriadis}, G., {Foley}, R.~J., {Rest}, A., {et~al.} 2019, \apjl, 870, L1,
  \dodoi{10.3847/2041-8213/aaedb0}

\bibitem[{{Edwards} {et~al.}(2012){Edwards}, {Pagnotta}, \& {Schaefer}}]{edw12}
{Edwards}, Z.~I., {Pagnotta}, A., \& {Schaefer}, B.~E. 2012, \apjl, 747, L19,
  \dodoi{10.1088/2041-8205/747/2/L19}

\bibitem[{{Fan} {et~al.}(2015){Fan}, {Bai}, {Zhang}, {Wang}, {Chang}, {Xin}, \&
  {Zhang}}]{LJT}
{Fan}, Y.-F., {Bai}, J.-M., {Zhang}, J.-J., {et~al.} 2015, Research in
  Astronomy and Astrophysics, 15, 918, \dodoi{10.1088/1674-4527/15/6/014}

\bibitem[{{Fan} {et~al.}(2016){Fan}, {Wang}, {Jiang}, {Wu}, {Li}, {Huang},
  {Xu}, {Hu}, {Zhu}, {Wang}, {Komossa}, \& {Zhang}}]{XLT216}
{Fan}, Z., {Wang}, H., {Jiang}, X., {et~al.} 2016, \pasp, 128, 115005,
  \dodoi{10.1088/1538-3873/128/969/115005}

\bibitem[{{Filippenko}(1997)}]{fili97}
{Filippenko}, A.~V. 1997, \araa, 35, 309,
  \dodoi{10.1146/annurev.astro.35.1.309}

\bibitem[{{Filippenko} {et~al.}(1992{\natexlab{a}}){Filippenko}, {Richmond},
  {Matheson}, {Shields}, {Burbidge}, {Cohen}, {Dickinson}, {Malkan}, {Nelson},
  {Pietz}, {Schlegel}, {Schmeer}, {Spinrad}, {Steidel}, {Tran}, \&
  {Wren}}]{fili92a}
{Filippenko}, A.~V., {Richmond}, M.~W., {Matheson}, T., {et~al.}
  1992{\natexlab{a}}, \apjl, 384, L15, \dodoi{10.1086/186252}

\bibitem[{{Filippenko} {et~al.}(1992{\natexlab{b}}){Filippenko}, {Richmond},
  {Branch}, {Gaskell}, {Herbst}, {Ford}, {Treffers}, {Matheson}, {Ho}, {Dey},
  {Sargent}, {Small}, \& {van Breugel}}]{fili92b}
{Filippenko}, A.~V., {Richmond}, M.~W., {Branch}, D., {et~al.}
  1992{\natexlab{b}}, \aj, 104, 1543, \dodoi{10.1086/116339}

\bibitem[{{Fink} {et~al.}(2010){Fink}, {R{\"o}pke}, {Hillebrandt},
  {Seitenzahl}, {Sim}, \& {Kromer}}]{fin10}
{Fink}, M., {R{\"o}pke}, F.~K., {Hillebrandt}, W., {et~al.} 2010, \aap, 514,
  A53, \dodoi{10.1051/0004-6361/200913892}

\bibitem[{{Folatelli} {et~al.}(2013){Folatelli}, {Morrell}, {Phillips},
  {Hsiao}, {Campillay}, {Contreras}, {Castell{\'o}n}, {Hamuy}, {Krzeminski},
  {Roth}, {Stritzinger}, {Burns}, {Freedman}, {Madore}, {Murphy}, {Persson},
  {Prieto}, {Suntzeff}, {Krisciunas}, {Anderson}, {F{\"o}rster}, {Maza},
  {Pignata}, {Rojas}, {Boldt}, {Salgado}, {Wyatt}, {Olivares E.}, {Gal-Yam}, \&
  {Sako}}]{fol13}
{Folatelli}, G., {Morrell}, N., {Phillips}, M.~M., {et~al.} 2013, \apj, 773,
  53, \dodoi{10.1088/0004-637X/773/1/53}

\bibitem[{{Freedman} {et~al.}(2009){Freedman}, {Burns}, {Phillips}, {Wyatt},
  {Persson}, {Madore}, {Contreras}, {Folatelli}, {Gonzalez}, {Hamuy}, {Hsiao},
  {Kelson}, {Morrell}, {Murphy}, {Roth}, {Stritzinger}, {Sturch}, {Suntzeff},
  {Astier}, {Balland}, {Bassett}, {Boldt}, {Carlberg}, {Conley}, {Frieman},
  {Garnavich}, {Guy}, {Hardin}, {Howell}, {Kessler}, {Lampeitl}, {Marriner},
  {Pain}, {Perrett}, {Regnault}, {Riess}, {Sako}, {Schneider}, {Sullivan}, \&
  {Wood-Vasey}}]{fre09}
{Freedman}, W.~L., {Burns}, C.~R., {Phillips}, M.~M., {et~al.} 2009, \apj, 704,
  1036, \dodoi{10.1088/0004-637X/704/2/1036}

\bibitem[{{Galbany} {et~al.}(2019){Galbany}, {Ashall}, {H{\"o}flich},
  {Gonz{\'a}lez-Gait{\'a}n}, {Taubenberger}, {Stritzinger}, {Hsiao}, {Mazzali},
  {Baron}, {Blondin}, {Bose}, {Bulla}, {Burke}, {Burns}, {Cartier}, {Chen},
  {Della Valle}, {Diamond}, {Guti{\'e}rrez}, {Harmanen}, {Hiramatsu},
  {Holoien}, {Hosseinzadeh}, {Howell}, {Huang}, {Inserra}, {de Jaeger}, {Jha},
  {Kangas}, {Kromer}, {Lyman}, {Maguire}, {Marion}, {Milisavljevic},
  {Prentice}, {Razza}, {Reynolds}, {Sand}, {Shappee}, {Shekhar}, {Smartt},
  {Stassun}, {Sullivan}, {Valenti}, {Villanueva}, {Wang}, {Wheeler}, {Zhai}, \&
  {Zhang}}]{gal19}
{Galbany}, L., {Ashall}, C., {H{\"o}flich}, P., {et~al.} 2019, \aap, 630, A76,
  \dodoi{10.1051/0004-6361/201935537}

\bibitem[{{Gall} {et~al.}(2018){Gall}, {Stritzinger}, {Ashall}, {Baron},
  {Burns}, {Hoeflich}, {Hsiao}, {Mazzali}, {Phillips}, {Filippenko},
  {Anderson}, {Benetti}, {Brown}, {Campillay}, {Challis}, {Contreras}, {Elias
  de la Rosa}, {Folatelli}, {Foley}, {Fraser}, {Holmbo}, {Marion}, {Morrell},
  {Pan}, {Pignata}, {Suntzeff}, {Taddia}, {Torres Robledo}, \&
  {Valenti}}]{gal18}
{Gall}, C., {Stritzinger}, M.~D., {Ashall}, C., {et~al.} 2018, \aap, 611, A58,
  \dodoi{10.1051/0004-6361/201730886}

\bibitem[{{Ganeshalingam} {et~al.}(2010){Ganeshalingam}, {Li}, {Filippenko},
  {Anderson}, {Foster}, {Gates}, {Griffith}, {Grigsby}, {Joubert}, {Leja},
  {Lowe}, {Macomber}, {Pritchard}, {Thrasher}, \& {Winslow}}]{gane10}
{Ganeshalingam}, M., {Li}, W., {Filippenko}, A.~V., {et~al.} 2010, \apjs, 190,
  418, \dodoi{10.1088/0067-0049/190/2/418}

\bibitem[{{Garnavich} {et~al.}(2004){Garnavich}, {Bonanos}, {Krisciunas},
  {Jha}, {Kirshner}, {Schlegel}, {Challis}, {Macri}, {Hatano}, {Branch},
  {Bothun}, \& {Freedman}}]{gar04}
{Garnavich}, P.~M., {Bonanos}, A.~Z., {Krisciunas}, K., {et~al.} 2004, \apj,
  613, 1120, \dodoi{10.1086/422986}

\bibitem[{{Gonz{\'a}lez-Gait{\'a}n} {et~al.}(2011){Gonz{\'a}lez-Gait{\'a}n},
  {Perrett}, {Sullivan}, {Conley}, {Howell}, {Carlberg}, {Astier}, {Balam},
  {Balland}, {Basa}, {Fouchez}, {Guy}, {Hardin}, {Hook.}, {Pain}, {Pritchet},
  {Regnault}, {Rich}, \& {Lidman}}]{gon11}
{Gonz{\'a}lez-Gait{\'a}n}, S., {Perrett}, K., {Sullivan}, M., {et~al.} 2011,
  \apj, 727, 107, \dodoi{10.1088/0004-637X/727/2/107}

\bibitem[{{Gonz{\'a}lez-Gait{\'a}n} {et~al.}(2012){Gonz{\'a}lez-Gait{\'a}n},
  {Conley}, {Bianco}, {Howell}, {Sullivan}, {Perrett}, {Carlberg}, {Astier},
  {Balam}, {Balland}, {Basa}, {Fouchez}, {Fourmanoit}, {Graham}, {Guy},
  {Hardin}, {Hook}, {Lidman}, {Pain}, {Palanque-Delabrouille}, {Pritchet},
  {Regnault}, {Rich}, \& {Ruhlmann-Kleider}}]{gon12}
{Gonz{\'a}lez-Gait{\'a}n}, S., {Conley}, A., {Bianco}, F.~B., {et~al.} 2012,
  \apj, 745, 44, \dodoi{10.1088/0004-637X/745/1/44}

\bibitem[{{Guillochon} {et~al.}(2017){Guillochon}, {Parrent}, {Kelley}, \&
  {Margutti}}]{snespace}
{Guillochon}, J., {Parrent}, J., {Kelley}, L.~Z., \& {Margutti}, R. 2017, {An
  Open Catalog for Supernova Data}, \dodoi{10.3847/1538-4357/835/1/64}

\bibitem[{{Guy} {et~al.}(2005){Guy}, {Astier}, {Nobili}, {Regnault}, \&
  {Pain}}]{guy05}
{Guy}, J., {Astier}, P., {Nobili}, S., {Regnault}, N., \& {Pain}, R. 2005,
  \aap, 443, 781, \dodoi{10.1051/0004-6361:20053025}

\bibitem[{{Hakobyan} {et~al.}(2016){Hakobyan}, {Karapetyan}, {Barkhudaryan},
  {Mamon}, {Kunth}, {Petrosian}, {Adibekyan}, {Aramyan}, \& {Turatto}}]{hak16}
{Hakobyan}, A.~A., {Karapetyan}, A.~G., {Barkhudaryan}, L.~V., {et~al.} 2016,
  \mnras, 456, 2848, \dodoi{10.1093/mnras/stv2853}

\bibitem[{{Hamuy} {et~al.}(1996){Hamuy}, {Phillips}, {Suntzeff}, {Schommer},
  {Maza}, \& {Aviles}}]{ham96a}
{Hamuy}, M., {Phillips}, M.~M., {Suntzeff}, N.~B., {et~al.} 1996, \aj, 112,
  2391, \dodoi{10.1086/118190}

\bibitem[{{Hamuy} {et~al.}(2002){Hamuy}, {Maza}, {Pinto}, {Phillips},
  {Suntzeff}, {Blum}, {Olsen}, {Pinfield}, {Ivanov}, {Augusteijn}, {Brillant},
  {Chadid}, {Cuby}, {Doublier}, {Hainaut}, {Le Floc'h}, {Lidman},
  {Petr-Gotzens}, {Pompei}, \& {Vanzi}}]{ham02}
{Hamuy}, M., {Maza}, J., {Pinto}, P.~A., {et~al.} 2002, \aj, 124, 417,
  \dodoi{10.1086/340968}

\bibitem[{{Hicken} {et~al.}(2009){Hicken}, {Challis}, {Jha}, {Kirshner},
  {Matheson}, {Modjaz}, {Rest}, {Wood-Vasey}, {Bakos}, {Barton}, {Berlind},
  {Bragg}, {Brice{\~n}o}, {Brown}, {Caldwell}, {Calkins}, {Cho}, {Ciupik},
  {Contreras}, {Dendy}, {Dosaj}, {Durham}, {Eriksen}, {Esquerdo}, {Everett},
  {Falco}, {Fernandez}, {Gaba}, {Garnavich}, {Graves}, {Green}, {Groner},
  {Hergenrother}, {Holman}, {Hradecky}, {Huchra}, {Hutchison}, {Jerius},
  {Jordan}, {Kilgard}, {Krauss}, {Luhman}, {Macri}, {Marrone}, {McDowell},
  {McIntosh}, {McNamara}, {Megeath}, {Mochejska}, {Munoz}, {Muzerolle},
  {Naranjo}, {Narayan}, {Pahre}, {Peters}, {Peterson}, {Rines}, {Ripman},
  {Roussanova}, {Schild}, {Sicilia-Aguilar}, {Sokoloski}, {Smalley}, {Smith},
  {Spahr}, {Stanek}, {Barmby}, {Blondin}, {Stubbs}, {Szentgyorgyi}, {Torres},
  {Vaz}, {Vikhlinin}, {Wang}, {Westover}, {Woods}, \& {Zhao}}]{hic09}
{Hicken}, M., {Challis}, P., {Jha}, S., {et~al.} 2009, \apj, 700, 331,
  \dodoi{10.1088/0004-637X/700/1/331}

\bibitem[{{H{\"o}flich} {et~al.}(2002){H{\"o}flich}, {Gerardy}, {Fesen}, \&
  {Sakai}}]{hof02}
{H{\"o}flich}, P., {Gerardy}, C.~L., {Fesen}, R.~A., \& {Sakai}, S. 2002, \apj,
  568, 791, \dodoi{10.1086/339063}

\bibitem[{{H{\"o}flich} \& {Khokhlov}(1996)}]{hoe96}
{H{\"o}flich}, P., \& {Khokhlov}, A. 1996, \apj, 457, 500,
  \dodoi{10.1086/176748}

\bibitem[{{H{\"o}flich} {et~al.}(2017){H{\"o}flich}, {Hsiao}, {Ashall},
  {Burns}, {Diamond}, {Phillips}, {Sand}, {Stritzinger}, {Suntzeff},
  {Contreras}, {Krisciunas}, {Morrell}, \& {Wang}}]{hoe17}
{H{\"o}flich}, P., {Hsiao}, E.~Y., {Ashall}, C., {et~al.} 2017, \apj, 846, 58,
  \dodoi{10.3847/1538-4357/aa84b2}

\bibitem[{{Hosseinzadeh} {et~al.}(2017){Hosseinzadeh}, {Sand}, {Valenti},
  {Brown}, {Howell}, {McCully}, {Kasen}, {Arcavi}, {Bostroem}, {Tartaglia},
  {Hsiao}, {Davis}, {Shahbandeh}, \& {Stritzinger}}]{hoss17}
{Hosseinzadeh}, G., {Sand}, D.~J., {Valenti}, S., {et~al.} 2017, \apjl, 845,
  L11, \dodoi{10.3847/2041-8213/aa8402}

\bibitem[{{Hough} {et~al.}(1987){Hough}, {Bailey}, {Rouse}, \&
  {Whittet}}]{hou87}
{Hough}, J.~H., {Bailey}, J.~A., {Rouse}, M.~F., \& {Whittet}, D.~C.~B. 1987,
  \mnras, 227, 1P, \dodoi{10.1093/mnras/227.1.1P}

\bibitem[{{Howell} {et~al.}(2001){Howell}, {H{\"o}flich}, {Wang}, \&
  {Wheeler}}]{how01}
{Howell}, D.~A., {H{\"o}flich}, P., {Wang}, L., \& {Wheeler}, J.~C. 2001, \apj,
  556, 302, \dodoi{10.1086/321584}

\bibitem[{{Hsiao} {et~al.}(2015){Hsiao}, {Burns}, {Contreras}, {H{\"o}flich},
  {Sand}, {Marion}, {Phillips}, {Stritzinger}, {Gonz{\'a}lez-Gait{\'a}n},
  {Mason}, {Folatelli}, {Parent}, {Gall}, {Amanullah}, {Anupama}, {Arcavi},
  {Banerjee}, {Beletsky}, {Blanc}, {Bloom}, {Brown}, {Campillay}, {Cao}, {De
  Cia}, {Diamond}, {Freedman}, {Gonzalez}, {Goobar}, {Holmbo}, {Howell},
  {Johansson}, {Kasliwal}, {Kirshner}, {Krisciunas}, {Kulkarni}, {Maguire},
  {Milne}, {Morrell}, {Nugent}, {Ofek}, {Osip}, {Palunas}, {Perley}, {Persson},
  {Piro}, {Rabus}, {Roth}, {Schiefelbein}, {Srivastav}, {Sullivan}, {Suntzeff},
  {Surace}, {Wo{\'z}niak}, \& {Yaron}}]{hsi15}
{Hsiao}, E.~Y., {Burns}, C.~R., {Contreras}, C., {et~al.} 2015, \aap, 578, A9,
  \dodoi{10.1051/0004-6361/201425297}

\bibitem[{{Hsiao} {et~al.}(2019){Hsiao}, {Phillips}, {Marion}, {Kirshner},
  {Morrell}, {Sand}, {Burns}, {Contreras}, {Hoeflich}, {Stritzinger},
  {Valenti}, {Anderson}, {Ashall}, {Baltay}, {Baron}, {Banerjee}, {Davis},
  {Diamond}, {Folatelli}, {Freedman}, {F{\"o}rster}, {Galbany}, {Gall},
  {Gonz{\'a}lez-Gait{\'a}n}, {Goobar}, {Hamuy}, {Holmbo}, {Kasliwal},
  {Krisciunas}, {Kumar}, {Lidman}, {Lu}, {Nugent}, {Perlmutter}, {Persson},
  {Piro}, {Rabinowitz}, {Roth}, {Ryder}, {Schmidt}, {Shahbandeh}, {Suntzeff},
  {Taddia}, {Uddin}, \& {Wang}}]{hsi19}
{Hsiao}, E.~Y., {Phillips}, M.~M., {Marion}, G.~H., {et~al.} 2019, \pasp, 131,
  014002, \dodoi{10.1088/1538-3873/aae961}

\bibitem[{{Huang} {et~al.}(2012){Huang}, {Li}, {Wang}, {Shang}, {Zhang}, {Hu},
  {Qiu}, \& {Jiang}}]{huang12}
{Huang}, F., {Li}, J.-Z., {Wang}, X.-F., {et~al.} 2012, Research in Astronomy
  and Astrophysics, 12, 1585, \dodoi{10.1088/1674-4527/12/11/012}

\bibitem[{{Hunter}(2007)}]{matplotlib}
{Hunter}, J.~D. 2007, Computing in Science and Engineering, 9, 90,
  \dodoi{10.1109/MCSE.2007.55}

\bibitem[{{Iben} \& {Tutukov}(1984)}]{iben84}
{Iben}, I., J., \& {Tutukov}, A.~V. 1984, \apjs, 54, 335,
  \dodoi{10.1086/190932}

\bibitem[{{Ivezi{\'c}} {et~al.}(2019){Ivezi{\'c}}, {Kahn}, {Tyson}, {Abel},
  {Acosta}, {Allsman}, {Alonso}, {AlSayyad}, {Anderson}, {Andrew}, {Angel},
  {Angeli}, {Ansari}, {Antilogus}, {Araujo}, {Armstrong}, {Arndt}, {Astier},
  {Aubourg}, {Auza}, {Axelrod}, {Bard}, {Barr}, {Barrau}, {Bartlett}, {Bauer},
  {Bauman}, {Baumont}, {Bechtol}, {Bechtol}, {Becker}, {Becla}, {Beldica},
  {Bellavia}, {Bianco}, {Biswas}, {Blanc}, {Blazek}, {Blandford}, {Bloom},
  {Bogart}, {Bond}, {Booth}, {Borgland}, {Borne}, {Bosch}, {Boutigny},
  {Brackett}, {Bradshaw}, {Brandt}, {Brown}, {Bullock}, {Burchat}, {Burke},
  {Cagnoli}, {Calabrese}, {Callahan}, {Callen}, {Carlin}, {Carlson},
  {Chandrasekharan}, {Charles-Emerson}, {Chesley}, {Cheu}, {Chiang}, {Chiang},
  {Chirino}, {Chow}, {Ciardi}, {Claver}, {Cohen-Tanugi}, {Cockrum}, {Coles},
  {Connolly}, {Cook}, {Cooray}, {Covey}, {Cribbs}, {Cui}, {Cutri}, {Daly},
  {Daniel}, {Daruich}, {Daubard}, {Daues}, {Dawson}, {Delgado}, {Dellapenna},
  {de Peyster}, {de Val-Borro}, {Digel}, {Doherty}, {Dubois},
  {Dubois-Felsmann}, {Durech}, {Economou}, {Eifler}, {Eracleous}, {Emmons},
  {Fausti Neto}, {Ferguson}, {Figueroa}, {Fisher-Levine}, {Focke}, {Foss},
  {Frank}, {Freemon}, {Gangler}, {Gawiser}, {Geary}, {Gee}, {Geha}, {Gessner},
  {Gibson}, {Gilmore}, {Glanzman}, {Glick}, {Goldina}, {Goldstein}, {Goodenow},
  {Graham}, {Gressler}, {Gris}, {Guy}, {Guyonnet}, {Haller}, {Harris},
  {Hascall}, {Haupt}, {Hernandez}, {Herrmann}, {Hileman}, {Hoblitt}, {Hodgson},
  {Hogan}, {Howard}, {Huang}, {Huffer}, {Ingraham}, {Innes}, {Jacoby}, {Jain},
  {Jammes}, {Jee}, {Jenness}, {Jernigan}, {Jevremovi{\'c}}, {Johns}, {Johnson},
  {Johnson}, {Jones}, {Juramy-Gilles}, {Juri{\'c}}, {Kalirai}, {Kallivayalil},
  {Kalmbach}, {Kantor}, {Karst}, {Kasliwal}, {Kelly}, {Kessler}, {Kinnison},
  {Kirkby}, {Knox}, {Kotov}, {Krabbendam}, {Krughoff}, {Kub{\'a}nek},
  {Kuczewski}, {Kulkarni}, {Ku}, {Kurita}, {Lage}, {Lambert}, {Lange},
  {Langton}, {Le Guillou}, {Levine}, {Liang}, {Lim}, {Lintott}, {Long},
  {Lopez}, {Lotz}, {Lupton}, {Lust}, {MacArthur}, {Mahabal}, {Mandelbaum},
  {Markiewicz}, {Marsh}, {Marshall}, {Marshall}, {May}, {McKercher}, {McQueen},
  {Meyers}, {Migliore}, {Miller}, {Mills}, {Miraval}, {Moeyens}, {Moolekamp},
  {Monet}, {Moniez}, {Monkewitz}, {Montgomery}, {Morrison}, {Mueller},
  {Muller}, {Mu{\~n}oz Arancibia}, {Neill}, {Newbry}, {Nief}, {Nomerotski},
  {Nordby}, {O'Connor}, {Oliver}, {Olivier}, {Olsen}, {O'Mullane}, {Ortiz},
  {Osier}, {Owen}, {Pain}, {Palecek}, {Parejko}, {Parsons}, {Pease},
  {Peterson}, {Peterson}, {Petravick}, {Libby Petrick}, {Petry},
  {Pierfederici}, {Pietrowicz}, {Pike}, {Pinto}, {Plante}, {Plate}, {Plutchak},
  {Price}, {Prouza}, {Radeka}, {Rajagopal}, {Rasmussen}, {Regnault}, {Reil},
  {Reiss}, {Reuter}, {Ridgway}, {Riot}, {Ritz}, {Robinson}, {Roby}, {Roodman},
  {Rosing}, {Roucelle}, {Rumore}, {Russo}, {Saha}, {Sassolas}, {Schalk},
  {Schellart}, {Schindler}, {Schmidt}, {Schneider}, {Schneider}, {Schoening},
  {Schumacher}, {Schwamb}, {Sebag}, {Selvy}, {Sembroski}, {Seppala}, {Serio},
  {Serrano}, {Shaw}, {Shipsey}, {Sick}, {Silvestri}, {Slater}, {Smith},
  {Smith}, {Sobhani}, {Soldahl}, {Storrie-Lombardi}, {Stover}, {Strauss},
  {Street}, {Stubbs}, {Sullivan}, {Sweeney}, {Swinbank}, {Szalay}, {Takacs},
  {Tether}, {Thaler}, {Thayer}, {Thomas}, {Thornton}, {Thukral}, {Tice},
  {Trilling}, {Turri}, {Van Berg}, {Vanden Berk}, {Vetter}, {Virieux},
  {Vucina}, {Wahl}, {Walkowicz}, {Walsh}, {Walter}, {Wang}, {Wang}, {Warner},
  {Wiecha}, {Willman}, {Winters}, {Wittman}, {Wolff}, {Wood-Vasey}, {Wu},
  {Xin}, {Yoachim}, \& {Zhan}}]{LSST}
{Ivezi{\'c}}, {\v{Z}}., {Kahn}, S.~M., {Tyson}, J.~A., {et~al.} 2019, \apj,
  873, 111, \dodoi{10.3847/1538-4357/ab042c}

\bibitem[{{Iwamoto} {et~al.}(1999){Iwamoto}, {Brachwitz}, {Nomoto},
  {Kishimoto}, {Umeda}, {Hix}, \& {Thielemann}}]{iwa99}
{Iwamoto}, K., {Brachwitz}, F., {Nomoto}, K., {et~al.} 1999, \apjs, 125, 439,
  \dodoi{10.1086/313278}

\bibitem[{{Jha} {et~al.}(2007){Jha}, {Riess}, \& {Kirshner}}]{jha07}
{Jha}, S., {Riess}, A.~G., \& {Kirshner}, R.~P. 2007, \apj, 659, 122,
  \dodoi{10.1086/512054}

\bibitem[{{Kasen}(2006)}]{kas06}
{Kasen}, D. 2006, \apj, 649, 939, \dodoi{10.1086/506588}

\bibitem[{{Khokhlov}(1991)}]{kho91}
{Khokhlov}, A.~M. 1991, \aap, 245, 114

\bibitem[{{Krisciunas} {et~al.}(2001){Krisciunas}, {Phillips}, {Stubbs},
  {Rest}, {Miknaitis}, {Riess}, {Suntzeff}, {Roth}, {Persson}, \&
  {Freedman}}]{kri01}
{Krisciunas}, K., {Phillips}, M.~M., {Stubbs}, C., {et~al.} 2001, \aj, 122,
  1616, \dodoi{10.1086/322120}

\bibitem[{{Krisciunas} {et~al.}(2009){Krisciunas}, {Marion}, {Suntzeff},
  {Blanc}, {Bufano}, {Candia}, {Cartier}, {Elias-Rosa}, {Espinoza}, {Gonzalez},
  {Gonzalez}, {Gonzalez}, {Gooding}, {Hamuy}, {Knox}, {Milne}, {Morrell},
  {Phillips}, {Stritzinger}, \& {Thomas-Osip}}]{kri09}
{Krisciunas}, K., {Marion}, G.~H., {Suntzeff}, N.~B., {et~al.} 2009, \aj, 138,
  1584, \dodoi{10.1088/0004-6256/138/6/1584}

\bibitem[{{Krisciunas} {et~al.}(2017){Krisciunas}, {Contreras}, {Burns},
  {Phillips}, {Stritzinger}, {Morrell}, {Hamuy}, {Anais}, {Boldt}, {Busta},
  {Campillay}, {Castell{\'o}n}, {Folatelli}, {Freedman}, {Gonz{\'a}lez},
  {Hsiao}, {Krzeminski}, {Persson}, {Roth}, {Salgado}, {Ser{\'o}n}, {Suntzeff},
  {Torres}, {Filippenko}, {Li}, {Madore}, {DePoy}, {Marshall}, {Rheault}, \&
  {Villanueva}}]{kri17}
{Krisciunas}, K., {Contreras}, C., {Burns}, C.~R., {et~al.} 2017, \aj, 154,
  211, \dodoi{10.3847/1538-3881/aa8df0}

\bibitem[{{Leibundgut} {et~al.}(1993){Leibundgut}, {Kirshner}, {Phillips},
  {Wells}, {Suntzeff}, {Hamuy}, {Schommer}, {Walker}, {Gonzalez}, {Ugarte},
  {Williams}, {Williger}, {Gomez}, {Marzke}, {Schmidt}, {Whitney}, {Caldwell},
  {Peters}, {Chaffee}, {Foltz}, {Rehner}, {Siciliano}, {Barnes}, {Cheng},
  {Hintzen}, {Kim}, {Maza}, {Parker}, {Porter}, {Schmidtke}, \&
  {Sonneborn}}]{lei93}
{Leibundgut}, B., {Kirshner}, R.~P., {Phillips}, M.~M., {et~al.} 1993, \aj,
  105, 301, \dodoi{10.1086/116427}

\bibitem[{{Li} {et~al.}(2003){Li}, {Filippenko}, {Chornock}, {Berger},
  {Berlind}, {Calkins}, {Challis}, {Fassnacht}, {Jha}, {Kirshner}, {Matheson},
  {Sargent}, {Simcoe}, {Smith}, \& {Squires}}]{liw03}
{Li}, W., {Filippenko}, A.~V., {Chornock}, R., {et~al.} 2003, \pasp, 115, 453,
  \dodoi{10.1086/374200}

\bibitem[{{Li} {et~al.}(2011{\natexlab{a}}){Li}, {Bloom}, {Podsiadlowski},
  {Miller}, {Cenko}, {Jha}, {Sullivan}, {Howell}, {Nugent}, {Butler}, {Ofek},
  {Kasliwal}, {Richards}, {Stockton}, {Shih}, {Bildsten}, {Shara}, {Bibby},
  {Filippenko}, {Ganeshalingam}, {Silverman}, {Kulkarni}, {Law}, {Poznanski},
  {Quimby}, {McCully}, {Patel}, {Maguire}, \& {Shen}}]{liw11b}
{Li}, W., {Bloom}, J.~S., {Podsiadlowski}, P., {et~al.} 2011{\natexlab{a}},
  \nat, 480, 348, \dodoi{10.1038/nature10646}

\bibitem[{{Li} {et~al.}(2011{\natexlab{b}}){Li}, {Leaman}, {Chornock},
  {Filippenko}, {Poznanski}, {Ganeshalingam}, {Wang}, {Modjaz}, {Jha}, {Foley},
  \& {Smith}}]{liw11a}
{Li}, W., {Leaman}, J., {Chornock}, R., {et~al.} 2011{\natexlab{b}}, \mnras,
  412, 1441, \dodoi{10.1111/j.1365-2966.2011.18160.x}

\bibitem[{{Li} {et~al.}(2019){Li}, {Wang}, {Vink{\'o}}, {Mo}, {Hosseinzadeh},
  {Sand}, {Zhang}, {Lin}, {PTSS/TNTS}, {Zhang}, {Wang}, {Zhang}, {Chen},
  {Xiang}, {Rui}, {Huang}, {Li}, {Zhang}, {Li}, {Baron}, {Derkacy}, {Zhao},
  {Sai}, {Zhang}, {Wang}, {LCO}, {Howell}, {McCully}, {Arcavi}, {Valenti},
  {Hiramatsu}, {Burke}, {KEGS}, {Rest}, {Garnavich}, {Tucker}, {Narayan},
  {Shaya}, {Margheim}, {Zenteno}, {Villar}, {UCSC}, {Dimitriadis}, {Foley},
  {Pan}, {Coulter}, {Fox}, {Jha}, {Jones}, {Kasen}, {Kilpatrick}, {Piro},
  {Riess}, {Rojas-Bravo}, {ASAS-SN}, {Shappee}, {Holoien}, {Stanek}, {Drout},
  {Auchettl}, {Kochanek}, {Brown}, {Bose}, {Bersier}, {Brimacombe}, {Chen},
  {Dong}, {Holmbo}, {Mu{\~n}oz}, {Mutel}, {Post}, {Prieto}, {Shields},
  {Tallon}, {Thompson}, {Vallely}, {Villanueva}, {Pan-STARRS}, {Smartt},
  {Smith}, {Chambers}, {Flewelling}, {Huber}, {Magnier}, {Waters}, {Schultz},
  {Bulger}, {Lowe}, {Willman}, {Konkoly/Texas}, {S{\'a}rneczky}, {P{\'a}l},
  {Wheeler}, {B{\'o}di}, {Bogn{\'a}r}, {Cs{\'a}k}, {Cseh}, {Cs{\"o}rnyei},
  {Hanyecz}, {Ign{\'a}cz}, {Kalup}, {K{\"o}nyves-T{\'o}th}, {Kriskovics},
  {Ordasi}, {Rajmon}, {S{\'o}dor}, {Szab{\'o}}, {Szak{\'a}ts}, {Zsidi},
  {Arizona}, {Milne}, {Andrews}, {Smith}, {Bilinski}, {Swift}, {Brown},
  {ePESSTO}, {Nordin}, {Williams}, {Galbany}, {Palmerio}, {Hook}, {Inserra},
  {Maguire}, {Cartier}, {Razza}, {Guti{\'e}rrez}, {North Carolina}, {Hermes},
  {Reding}, {Kaiser}, {ATLAS}, {Tonry}, {Heinze}, {Denneau}, {Weiland},
  {Stalder}, {K2 Mission Team}, {Barentsen}, {Dotson}, {Barclay},
  {Gully-Santiago}, {Hedges}, {Cody}, {Howell}, {Kepler Spacecraft Team},
  {Coughlin}, {Van Cleve}, {Cardoso}, {Larson}, {McCalmont-Everton},
  {Peterson}, {Ross}, {Reedy}, {Osborne}, {McGinn}, {Kohnert}, {Migliorini},
  {Wheaton}, {Spencer}, {Labonde}, {Castillo}, {Beerman}, {Steward}, {Hanley},
  {Larsen}, {Gangopadhyay}, {Kloetzel}, {Weschler}, {Nystrom}, {Moffatt},
  {Redick}, {Griest}, {Packard}, {Muszynski}, {Kampmeier}, {Bjella}, {Flynn},
  \& {Elsaesser}}]{liw19a}
{Li}, W., {Wang}, X., {Vink{\'o}}, J., {et~al.} 2019, \apj, 870, 12,
  \dodoi{10.3847/1538-4357/aaec74}

\bibitem[{{Li} {et~al.}(2021){Li}, {Wang}, {Bulla}, {Pan}, {Wang}, {Mo},
  {Zhang}, {Wu}, {Zhang}, {Zhang}, {Xiang}, {Lin}, {Sai}, {Zhang}, {Chen}, \&
  {Yan}}]{liw21}
{Li}, W., {Wang}, X., {Bulla}, M., {et~al.} 2021, \apj, 906, 99,
  \dodoi{10.3847/1538-4357/abc9b5}

\bibitem[{{Lu} {et~al.}(1993){Lu}, {Hoffman}, {Groff}, {Roos}, \&
  {Lamphier}}]{lu93}
{Lu}, N.~Y., {Hoffman}, G.~L., {Groff}, T., {Roos}, T., \& {Lamphier}, C. 1993,
  \apjs, 88, 383, \dodoi{10.1086/191826}

\bibitem[{Maeda \& Terada(2016)}]{mae16}
Maeda, K., \& Terada, Y. 2016, International Journal of Modern Physics D, 25,
  1630024, \dodoi{10.1142/s021827181630024x}

\bibitem[{{Maoz} {et~al.}(2014){Maoz}, {Mannucci}, \& {Nelemans}}]{maoz14}
{Maoz}, D., {Mannucci}, F., \& {Nelemans}, G. 2014, \araa, 52, 107,
  \dodoi{10.1146/annurev-astro-082812-141031}

\bibitem[{{Maraston} {et~al.}(2009){Maraston}, {Str{\"o}mb{\"a}ck}, {Thomas},
  {Wake}, \& {Nichol}}]{mar09}
{Maraston}, C., {Str{\"o}mb{\"a}ck}, G., {Thomas}, D., {Wake}, D.~A., \&
  {Nichol}, R.~C. 2009, \mnras, 394, L107,
  \dodoi{10.1111/j.1745-3933.2009.00621.x}

\bibitem[{{Mazzali} {et~al.}(1997){Mazzali}, {Chugai}, {Turatto}, {Lucy},
  {Danziger}, {Cappellaro}, {della Valle}, \& {Benetti}}]{maz97}
{Mazzali}, P.~A., {Chugai}, N., {Turatto}, M., {et~al.} 1997, \mnras, 284, 151,
  \dodoi{10.1093/mnras/284.1.151}

\bibitem[{{Modjaz} {et~al.}(2001){Modjaz}, {Li}, {Filippenko}, {King},
  {Leonard}, {Matheson}, {Treffers}, \& {Riess}}]{mod01}
{Modjaz}, M., {Li}, W., {Filippenko}, A.~V., {et~al.} 2001, \pasp, 113, 308,
  \dodoi{10.1086/319338}

\bibitem[{{NASA/IPAC Extragalactic Database (NED)}(2019)}]{NEDGWF}
{NASA/IPAC Extragalactic Database (NED)}. 2019, NED Gravitational Wave
  Follow-up (GWF) Service,  IPAC, \dodoi{10.26132/NED3}

\bibitem[{{Nugent} {et~al.}(1995){Nugent}, {Phillips}, {Baron}, {Branch}, \&
  {Hauschildt}}]{nug95}
{Nugent}, P., {Phillips}, M., {Baron}, E., {Branch}, D., \& {Hauschildt}, P.
  1995, \apjl, 455, L147, \dodoi{10.1086/309846}

\bibitem[{{O'Donnell}(1994)}]{odo94}
{O'Donnell}, J.~E. 1994, \apj, 422, 158, \dodoi{10.1086/173713}

\bibitem[{{Oliphant}(2007)}]{scipy}
{Oliphant}, T.~E. 2007, Computing in Science and Engineering, 9, 10,
  \dodoi{10.1109/MCSE.2007.58}

\bibitem[{{Pakmor} {et~al.}(2010){Pakmor}, {Kromer}, {R{\"o}pke}, {Sim},
  {Ruiter}, \& {Hillebrandt}}]{pak10}
{Pakmor}, R., {Kromer}, M., {R{\"o}pke}, F.~K., {et~al.} 2010, \nat, 463, 61,
  \dodoi{10.1038/nature08642}

\bibitem[{{Pakmor} {et~al.}(2012){Pakmor}, {Kromer}, {Taubenberger}, {Sim},
  {R{\"o}pke}, \& {Hillebrandt}}]{pak12}
{Pakmor}, R., {Kromer}, M., {Taubenberger}, S., {et~al.} 2012, \apjl, 747, L10,
  \dodoi{10.1088/2041-8205/747/1/L10}

\bibitem[{{Pakmor} {et~al.}(2013){Pakmor}, {Kromer}, {Taubenberger}, \&
  {Springel}}]{pak13}
{Pakmor}, R., {Kromer}, M., {Taubenberger}, S., \& {Springel}, V. 2013, \apjl,
  770, L8, \dodoi{10.1088/2041-8205/770/1/L8}

\bibitem[{{Papadogiannakis} {et~al.}(2019){Papadogiannakis}, {Dhawan},
  {Morosin}, \& {Goobar}}]{pap19}
{Papadogiannakis}, S., {Dhawan}, S., {Morosin}, R., \& {Goobar}, A. 2019,
  \mnras, 485, 2343, \dodoi{10.1093/mnras/stz493}

\bibitem[{{Pastorello} {et~al.}(2007){Pastorello}, {Mazzali}, {Pignata},
  {Benetti}, {Cappellaro}, {Filippenko}, {Li}, {Meikle}, {Arkharov}, {Blanc},
  {Bufano}, {Derekas}, {Dolci}, {Elias-Rosa}, {Foley}, {Ganeshalingam},
  {Harutyunyan}, {Kiss}, {Kotak}, {Larionov}, {Lucey}, {Napoleone},
  {Navasardyan}, {Patat}, {Rich}, {Ryder}, {Salvo}, {Schmidt}, {Stanishev},
  {Sz{\'e}kely}, {Taubenberger}, {Temporin}, {Turatto}, \&
  {Hillebrandt}}]{pas07b}
{Pastorello}, A., {Mazzali}, P.~A., {Pignata}, G., {et~al.} 2007, \mnras, 377,
  1531, \dodoi{10.1111/j.1365-2966.2007.11700.x}

\bibitem[{{Patat} {et~al.}(2007){Patat}, {Chandra}, {Chevalier}, {Justham},
  {Podsiadlowski}, {Wolf}, {Gal-Yam}, {Pasquini}, {Crawford}, {Mazzali},
  {Pauldrach}, {Nomoto}, {Benetti}, {Cappellaro}, {Elias-Rosa}, {Hillebrandt},
  {Leonard}, {Pastorello}, {Renzini}, {Sabbadin}, {Simon}, \&
  {Turatto}}]{pat07}
{Patat}, F., {Chandra}, P., {Chevalier}, R., {et~al.} 2007, Science, 317, 924,
  \dodoi{10.1126/science.1143005}

\bibitem[{{Perlmutter} {et~al.}(1999){Perlmutter}, {Aldering}, {Goldhaber},
  {Knop}, {Nugent}, {Castro}, {Deustua}, {Fabbro}, {Goobar}, {Groom}, {Hook},
  {Kim}, {Kim}, {Lee}, {Nunes}, {Pain}, {Pennypacker}, {Quimby}, {Lidman},
  {Ellis}, {Irwin}, {McMahon}, {Ruiz-Lapuente}, {Walton}, {Schaefer}, {Boyle},
  {Filippenko}, {Matheson}, {Fruchter}, {Panagia}, {Newberg}, {Couch}, \&
  {Project}}]{per99}
{Perlmutter}, S., {Aldering}, G., {Goldhaber}, G., {et~al.} 1999, \apj, 517,
  565, \dodoi{10.1086/307221}

\bibitem[{{Phillips}(1993)}]{phi93}
{Phillips}, M.~M. 1993, \apjl, 413, L105, \dodoi{10.1086/186970}

\bibitem[{{Phillips} {et~al.}(1999){Phillips}, {Lira}, {Suntzeff}, {Schommer},
  {Hamuy}, \& {Maza}}]{phi99}
{Phillips}, M.~M., {Lira}, P., {Suntzeff}, N.~B., {et~al.} 1999, \aj, 118,
  1766, \dodoi{10.1086/301032}

\bibitem[{{Phillips} {et~al.}(1992){Phillips}, {Wells}, {Suntzeff}, {Hamuy},
  {Leibundgut}, {Kirshner}, \& {Foltz}}]{phi92}
{Phillips}, M.~M., {Wells}, L.~A., {Suntzeff}, N.~B., {et~al.} 1992, \aj, 103,
  1632, \dodoi{10.1086/116177}

\bibitem[{{Phillips} {et~al.}(1987){Phillips}, {Phillips}, {Heathcote},
  {Blanco}, {Geisler}, {Hamilton}, {Suntzeff}, {Jablonski}, {Steiner},
  {Cowley}, {Schmidtke}, {Wyckoff}, {Hutchings}, {Tonry}, {Strauss},
  {Thorstensen}, {Honey}, {Maza}, {Ruiz}, {Landolt}, {Uomoto}, {Rich},
  {Grindlay}, {Cohn}, {Smith}, {Lutz}, {Lavery}, \& {Saha}}]{phi87}
{Phillips}, M.~M., {Phillips}, A.~C., {Heathcote}, S.~R., {et~al.} 1987, \pasp,
  99, 592, \dodoi{10.1086/132020}

\bibitem[{{Phillips} {et~al.}(2019){Phillips}, {Contreras}, {Hsiao}, {Morrell},
  {Burns}, {Stritzinger}, {Ashall}, {Freedman}, {Hoeflich}, {Persson}, {Piro},
  {Suntzeff}, {Uddin}, {Anais}, {Baron}, {Busta}, {Campillay}, {Castell{\'o}n},
  {Corco}, {Diamond}, {Gall}, {Gonzalez}, {Holmbo}, {Krisciunas}, {Roth},
  {Ser{\'o}n}, {Taddia}, {Torres}, {Anderson}, {Baltay}, {Folatelli},
  {Galbany}, {Goobar}, {Hadjiyska}, {Hamuy}, {Kasliwal}, {Lidman}, {Nugent},
  {Perlmutter}, {Rabinowitz}, {Ryder}, {Schmidt}, {Shappee}, \&
  {Walker}}]{phi19}
{Phillips}, M.~M., {Contreras}, C., {Hsiao}, E.~Y., {et~al.} 2019, \pasp, 131,
  014001, \dodoi{10.1088/1538-3873/aae8bd}

\bibitem[{{Polin} {et~al.}(2019){Polin}, {Nugent}, \& {Kasen}}]{pol19}
{Polin}, A., {Nugent}, P., \& {Kasen}, D. 2019, \apj, 873, 84,
  \dodoi{10.3847/1538-4357/aafb6a}

\bibitem[{{Riess} {et~al.}(1999){Riess}, {Filippenko}, {Li}, \&
  {Schmidt}}]{riess99}
{Riess}, A.~G., {Filippenko}, A.~V., {Li}, W., \& {Schmidt}, B.~P. 1999, \aj,
  118, 2668, \dodoi{10.1086/301144}

\bibitem[{{Riess} {et~al.}(1998){Riess}, {Filippenko}, {Challis},
  {Clocchiatti}, {Diercks}, {Garnavich}, {Gilliland}, {Hogan}, {Jha},
  {Kirshner}, {Leibundgut}, {Phillips}, {Reiss}, {Schmidt}, {Schommer},
  {Smith}, {Spyromilio}, {Stubbs}, {Suntzeff}, \& {Tonry}}]{riess98}
{Riess}, A.~G., {Filippenko}, A.~V., {Challis}, P., {et~al.} 1998, \aj, 116,
  1009, \dodoi{10.1086/300499}

\bibitem[{{Riess} {et~al.}(2007){Riess}, {Strolger}, {Casertano}, {Ferguson},
  {Mobasher}, {Gold}, {Challis}, {Filippenko}, {Jha}, {Li}, {Tonry}, {Foley},
  {Kirshner}, {Dickinson}, {MacDonald}, {Eisenstein}, {Livio}, {Younger}, {Xu},
  {Dahl{\'e}n}, \& {Stern}}]{riess07}
{Riess}, A.~G., {Strolger}, L.-G., {Casertano}, S., {et~al.} 2007, \apj, 659,
  98, \dodoi{10.1086/510378}

\bibitem[{{Riess} {et~al.}(2016){Riess}, {Macri}, {Hoffmann}, {Scolnic},
  {Casertano}, {Filippenko}, {Tucker}, {Reid}, {Jones}, {Silverman},
  {Chornock}, {Challis}, {Yuan}, {Brown}, \& {Foley}}]{riess16}
{Riess}, A.~G., {Macri}, L.~M., {Hoffmann}, S.~L., {et~al.} 2016, \apj, 826,
  56, \dodoi{10.3847/0004-637X/826/1/56}

\bibitem[{{R{\"o}ser} {et~al.}(2008){R{\"o}ser}, {Schilbach}, {Schwan},
  {Kharchenko}, {Piskunov}, \& {Scholz}}]{ros08}
{R{\"o}ser}, S., {Schilbach}, E., {Schwan}, H., {et~al.} 2008, \aap, 488, 401,
  \dodoi{10.1051/0004-6361:200809775}

\bibitem[{{Ruiz-Lapuente} {et~al.}(1992){Ruiz-Lapuente}, {Cappellaro},
  {Turatto}, {Gouiffes}, {Danziger}, {della Valle}, \& {Lucy}}]{ruiz92}
{Ruiz-Lapuente}, P., {Cappellaro}, E., {Turatto}, M., {et~al.} 1992, \apjl,
  387, L33, \dodoi{10.1086/186299}

\bibitem[{{Ruiz-Lapuente} {et~al.}(1993){Ruiz-Lapuente}, {Jeffery}, {Challis},
  {Filippenko}, {Kirshner}, {Ho}, {Schmidt}, {S{\'a}nchez}, \&
  {Canal}}]{ruiz93}
{Ruiz-Lapuente}, P., {Jeffery}, D.~J., {Challis}, P.~M., {et~al.} 1993, \nat,
  365, 728, \dodoi{10.1038/365728a0}

\bibitem[{{Sandage} {et~al.}(2006){Sandage}, {Tammann}, {Saha}, {Reindl},
  {Macchetto}, \& {Panagia}}]{sand06}
{Sandage}, A., {Tammann}, G.~A., {Saha}, A., {et~al.} 2006, \apj, 653, 843,
  \dodoi{10.1086/508853}

\bibitem[{{Santander-Garc{\'\i}a} {et~al.}(2015){Santander-Garc{\'\i}a},
  {Rodr{\'\i}guez-Gil}, {Corradi}, {Jones}, {Miszalski}, {Boffin},
  {Rubio-D{\'\i}ez}, \& {Kotze}}]{sant15}
{Santander-Garc{\'\i}a}, M., {Rodr{\'\i}guez-Gil}, P., {Corradi}, R.~L.~M.,
  {et~al.} 2015, \nat, 519, 63, \dodoi{10.1038/nature14124}

\bibitem[{{Sato} {et~al.}(2015){Sato}, {Nakasato}, {Tanikawa}, {Nomoto},
  {Maeda}, \& {Hachisu}}]{sat15}
{Sato}, Y., {Nakasato}, N., {Tanikawa}, A., {et~al.} 2015, \apj, 807, 105,
  \dodoi{10.1088/0004-637X/807/1/105}

\bibitem[{{Schaefer} \& {Pagnotta}(2012)}]{schae12}
{Schaefer}, B.~E., \& {Pagnotta}, A. 2012, \nat, 481, 164,
  \dodoi{10.1038/nature10692}

\bibitem[{{Schlafly} \& {Finkbeiner}(2011)}]{schla11}
{Schlafly}, E.~F., \& {Finkbeiner}, D.~P. 2011, \apj, 737, 103,
  \dodoi{10.1088/0004-637X/737/2/103}

\bibitem[{{Shappee} {et~al.}(2019){Shappee}, {Holoien}, {Drout}, {Auchettl},
  {Stritzinger}, {Kochanek}, {Stanek}, {Shaya}, {Narayan}, {ASAS-SN}, {Brown},
  {Bose}, {Bersier}, {Brimacombe}, {Chen}, {Dong}, {Holmbo}, {Katz},
  {Mu{\~n}oz}, {Mutel}, {Post}, {Prieto}, {Shields}, {Tallon}, {Thompson},
  {Vallely}, {Villanueva}, {ATLAS}, {Denneau}, {Flewelling}, {Heinze}, {Smith},
  {Stalder}, {Tonry}, {Weiland}, {Kepler/K2}, {Barclay}, {Barentsen}, {Cody},
  {Dotson}, {Foerster}, {Garnavich}, {Gully-Santiago}, {Hedges}, {Howell},
  {Kasen}, {Margheim}, {Mushotzky}, {Rest}, {Tucker}, {Villar}, {Zenteno},
  {Kepler Spacecraft Team}, {Beerman}, {Bjella}, {Castillo}, {Coughlin},
  {Elsaesser}, {Flynn}, {Gangopadhyay}, {Griest}, {Hanley}, {Kampmeier},
  {Kloetzel}, {Kohnert}, {Labonde}, {Larsen}, {Larson}, {McCalmont-Everton},
  {McGinn}, {Migliorini}, {Moffatt}, {Muszynski}, {Nystrom}, {Osborne},
  {Packard}, {Peterson}, {Redick}, {Reedy}, {Ross}, {Spencer}, {Steward}, {Van
  Cleve}, {Cardoso}, {Weschler}, {Wheaton}, {Pan-STARRS}, {Bulger}, {Chambers},
  {Flewelling}, {Huber}, {Lowe}, {Magnier}, {Schultz}, {Waters}, {Willman},
  {PTSS/TNTS}, {Baron}, {Chen}, {Derkacy}, {Huang}, {Li}, {Li}, {Li}, {Mo},
  {Rui}, {Sai}, {Wang}, {Wang}, {Wang}, {Xiang}, {Zhang}, {Zhang}, {Zhang},
  {Zhang}, {Zhang}, {Zhao}, {Brown}, {Hermes}, {Nordin}, {Points}, {S{\'o}dor},
  {Strampelli}, \& {Zenteno}}]{shappee19}
{Shappee}, B.~J., {Holoien}, T.~W.~S., {Drout}, M.~R., {et~al.} 2019, \apj,
  870, 13, \dodoi{10.3847/1538-4357/aaec79}

\bibitem[{{Silverman} {et~al.}(2012){Silverman}, {Foley}, {Filippenko},
  {Ganeshalingam}, {Barth}, {Chornock}, {Griffith}, {Kong}, {Lee}, {Leonard},
  {Matheson}, {Miller}, {Steele}, {Barris}, {Bloom}, {Cobb}, {Coil},
  {Desroches}, {Gates}, {Ho}, {Jha}, {Kandrashoff}, {Li}, {Mandel}, {Modjaz},
  {Moore}, {Mostardi}, {Papenkova}, {Park}, {Perley}, {Poznanski}, {Reuter},
  {Scala}, {Serduke}, {Shields}, {Swift}, {Tonry}, {Van Dyk}, {Wang}, \&
  {Wong}}]{silver12}
{Silverman}, J.~M., {Foley}, R.~J., {Filippenko}, A.~V., {et~al.} 2012, \mnras,
  425, 1789, \dodoi{10.1111/j.1365-2966.2012.21270.x}

\bibitem[{{Simcoe} {et~al.}(2013){Simcoe}, {Burgasser}, {Schechter}, {Fishner},
  {Bernstein}, {Bigelow}, {Pipher}, {Forrest}, {McMurtry}, {Smith}, \&
  {Bochanski}}]{sim13}
{Simcoe}, R.~A., {Burgasser}, A.~J., {Schechter}, P.~L., {et~al.} 2013, \pasp,
  125, 270, \dodoi{10.1086/670241}

\bibitem[{{Sternberg} {et~al.}(2011){Sternberg}, {Gal-Yam}, {Simon}, {Leonard},
  {Quimby}, {Phillips}, {Morrell}, {Thompson}, {Ivans}, {Marshall},
  {Filippenko}, {Marcy}, {Bloom}, {Patat}, {Foley}, {Yong}, {Penprase},
  {Beeler}, {Allende Prieto}, \& {Stringfellow}}]{ster11}
{Sternberg}, A., {Gal-Yam}, A., {Simon}, J.~D., {et~al.} 2011, Science, 333,
  856, \dodoi{10.1126/science.1203836}

\bibitem[{{Stetson}(1987)}]{stet87}
{Stetson}, P.~B. 1987, \pasp, 99, 191, \dodoi{10.1086/131977}

\bibitem[{Taubenberger(2017)}]{tau17}
Taubenberger, S. 2017, in Handbook of Supernovae (Springer International
  Publishing), 317--373, \dodoi{10.1007/978-3-319-21846-5_37}

\bibitem[{{Taubenberger} {et~al.}(2008){Taubenberger}, {Hachinger}, {Pignata},
  {Mazzali}, {Contreras}, {Valenti}, {Pastorello}, {Elias-Rosa},
  {B{\"a}rnbantner}, {Barwig}, {Benetti}, {Dolci}, {Fliri}, {Folatelli},
  {Freedman}, {Gonzalez}, {Hamuy}, {Krzeminski}, {Morrell}, {Navasardyan},
  {Persson}, {Phillips}, {Ries}, {Roth}, {Suntzeff}, {Turatto}, \&
  {Hillebrandt}}]{tau08}
{Taubenberger}, S., {Hachinger}, S., {Pignata}, G., {et~al.} 2008, \mnras, 385,
  75, \dodoi{10.1111/j.1365-2966.2008.12843.x}

\bibitem[{{Taubenberger} {et~al.}(2011){Taubenberger}, {Benetti}, {Childress},
  {Pakmor}, {Hachinger}, {Mazzali}, {Stanishev}, {Elias-Rosa}, {Agnoletto},
  {Bufano}, {Ergon}, {Harutyunyan}, {Inserra}, {Kankare}, {Kromer},
  {Navasardyan}, {Nicolas}, {Pastorello}, {Prosperi}, {Salgado}, {Sollerman},
  {Stritzinger}, {Turatto}, {Valenti}, \& {Hillebrandt}}]{tau11}
{Taubenberger}, S., {Benetti}, S., {Childress}, M., {et~al.} 2011, \mnras, 412,
  2735, \dodoi{10.1111/j.1365-2966.2010.18107.x}

\bibitem[{{Turatto} {et~al.}(1996){Turatto}, {Benetti}, {Cappellaro},
  {Danziger}, {Della Valle}, {Gouiffes}, {Mazzali}, \& {Patat}}]{tur96}
{Turatto}, M., {Benetti}, S., {Cappellaro}, E., {et~al.} 1996, \mnras, 283, 1,
  \dodoi{10.1093/mnras/283.1.1}

\bibitem[{{Turatto} {et~al.}(1998){Turatto}, {Piemonte}, {Benetti},
  {Cappellaro}, {Mazzali}, {Danziger}, \& {Patat}}]{tur98}
{Turatto}, M., {Piemonte}, A., {Benetti}, S., {et~al.} 1998, \aj, 116, 2431,
  \dodoi{10.1086/300622}

\bibitem[{{van der Walt} {et~al.}(2011){van der Walt}, {Colbert}, \&
  {Varoquaux}}]{numpy}
{van der Walt}, S., {Colbert}, S.~C., \& {Varoquaux}, G. 2011, Computing in
  Science and Engineering, 13, 22, \dodoi{10.1109/MCSE.2011.37}

\bibitem[{{Walker} {et~al.}(2015){Walker}, {Baltay}, {Campillay}, {Citrenbaum},
  {Contreras}, {Ellman}, {Feindt}, {Gonz{\'a}lez}, {Graham}, {Hadjiyska},
  {Hsiao}, {Krisciunas}, {McKinnon}, {Ment}, {Morrell}, {Nugent}, {Phillips},
  {Rabinowitz}, {Rostami}, {Ser{\'o}n}, {Stritzinger}, {Sullivan}, \&
  {Tucker}}]{wal15}
{Walker}, E.~S., {Baltay}, C., {Campillay}, A., {et~al.} 2015, \apjs, 219, 13,
  \dodoi{10.1088/0067-0049/219/1/13}

\bibitem[{{Wang}(2018)}]{wb18}
{Wang}, B. 2018, Research in Astronomy and Astrophysics, 18, 049,
  \dodoi{10.1088/1674-4527/18/5/49}

\bibitem[{{Wang} \& {Han}(2012)}]{wb12}
{Wang}, B., \& {Han}, Z. 2012, \nar, 56, 122,
  \dodoi{10.1016/j.newar.2012.04.001}

\bibitem[{{Wang} {et~al.}(2019){Wang}, {Bai}, {Fan}, {Mao}, {Chang}, {Xin},
  {Zhang}, {Lun}, {Wang}, {Zhang}, {Ying}, {Lu}, {Wang}, {Ji}, {Xiong}, {Yu},
  {Ding}, {Ye}, {Xing}, {Yi}, {Xu}, {Zheng}, {Feng}, {He}, {Wang}, {Liu},
  {Chen}, {Xu}, {Qin}, {Zhang}, {Tan}, {Li}, {Lou}, {Li}, \& {Liu}}]{YFOSC}
{Wang}, C.-J., {Bai}, J.-M., {Fan}, Y.-F., {et~al.} 2019, Research in Astronomy
  and Astrophysics, 19, 149, \dodoi{10.1088/1674-4527/19/10/149}

\bibitem[{{Wang} {et~al.}(2013{\natexlab{a}}){Wang}, {Chen}, {Zhang}, {Zhou},
  {Martignoni}, {Marion}, {Challis}, \& {Caulkins}}]{wxf13a}
{Wang}, X., {Chen}, J., {Zhang}, T., {et~al.} 2013{\natexlab{a}}, Central
  Bureau Electronic Telegrams, 3370, 1

\bibitem[{{Wang} {et~al.}(2013{\natexlab{b}}){Wang}, {Wang}, {Filippenko},
  {Zhang}, \& {Zhao}}]{wxf13b}
{Wang}, X., {Wang}, L., {Filippenko}, A.~V., {Zhang}, T., \& {Zhao}, X.
  2013{\natexlab{b}}, Science, 340, 170, \dodoi{10.1126/science.1231502}

\bibitem[{{Wang} {et~al.}(2006){Wang}, {Wang}, {Pain}, {Zhou}, \& {Li}}]{wxf06}
{Wang}, X., {Wang}, L., {Pain}, R., {Zhou}, X., \& {Li}, Z. 2006, \apj, 645,
  488, \dodoi{10.1086/504312}

\bibitem[{{Wang} {et~al.}(2005){Wang}, {Wang}, {Zhou}, {Lou}, \& {Li}}]{wxf05}
{Wang}, X., {Wang}, L., {Zhou}, X., {Lou}, Y.-Q., \& {Li}, Z. 2005, \apjl, 620,
  L87, \dodoi{10.1086/428774}

\bibitem[{{Wang} {et~al.}(2008){Wang}, {Li}, {Filippenko}, {Krisciunas},
  {Suntzeff}, {Li}, {Zhang}, {Deng}, {Foley}, {Ganeshalingam}, {Li}, {Lou},
  {Qiu}, {Shang}, {Silverman}, {Zhang}, \& {Zhang}}]{wxf08}
{Wang}, X., {Li}, W., {Filippenko}, A.~V., {et~al.} 2008, \apj, 675, 626,
  \dodoi{10.1086/526413}

\bibitem[{{Wang} {et~al.}(2009{\natexlab{a}}){Wang}, {Filippenko},
  {Ganeshalingam}, {Li}, {Silverman}, {Wang}, {Chornock}, {Foley}, {Gates},
  {Macomber}, {Serduke}, {Steele}, \& {Wong}}]{wxf09a}
{Wang}, X., {Filippenko}, A.~V., {Ganeshalingam}, M., {et~al.}
  2009{\natexlab{a}}, \apjl, 699, L139, \dodoi{10.1088/0004-637X/699/2/L139}

\bibitem[{{Wang} {et~al.}(2009{\natexlab{b}}){Wang}, {Li}, {Filippenko},
  {Foley}, {Kirshner}, {Modjaz}, {Bloom}, {Brown}, {Carter}, {Friedman},
  {Gal-Yam}, {Ganeshalingam}, {Hicken}, {Krisciunas}, {Milne}, {Silverman},
  {Suntzeff}, {Wood-Vasey}, {Cenko}, {Challis}, {Fox}, {Kirkman}, {Li}, {Li},
  {Malkan}, {Moore}, {Reitzel}, {Rich}, {Serduke}, {Shang}, {Steele}, {Swift},
  {Tao}, {Wong}, \& {Zhang}}]{wxf09b}
{Wang}, X., {Li}, W., {Filippenko}, A.~V., {et~al.} 2009{\natexlab{b}}, \apj,
  697, 380, \dodoi{10.1088/0004-637X/697/1/380}

\bibitem[{{Webbink}(1984)}]{web84}
{Webbink}, R.~F. 1984, \apj, 277, 355, \dodoi{10.1086/161701}

\bibitem[{{Whelan} \& {Iben}(1973)}]{whe73}
{Whelan}, J., \& {Iben}, Icko, J. 1973, \apj, 186, 1007, \dodoi{10.1086/152565}

\bibitem[{{Wilkinson} {et~al.}(2017){Wilkinson}, {Maraston}, {Goddard},
  {Thomas}, \& {Parikh}}]{wil17}
{Wilkinson}, D.~M., {Maraston}, C., {Goddard}, D., {Thomas}, D., \& {Parikh},
  T. 2017, \mnras, 472, 4297, \dodoi{10.1093/mnras/stx2215}

\bibitem[{{Woosley}(1990)}]{woo90}
{Woosley}, S.~E. 1990, in Supernovae, 182--212

\bibitem[{{Woosley} \& {Weaver}(1994)}]{woo94b}
{Woosley}, S.~E., \& {Weaver}, T.~A. 1994, \apj, 423, 371,
  \dodoi{10.1086/173813}

\bibitem[{{Wyatt} {et~al.}(2021){Wyatt}, {Sand}, {Hsiao}, {Burns}, {Valenti},
  {Bostroem}, {Lundquist}, {Galbany}, {Lu}, {Ashall}, {Diamond}, {Filippenko},
  {Graham}, {Hoeflich}, {Kirshner}, {Krisciunas}, {Marion}, {Morrell},
  {Persson}, {Phillips}, {Stritzinger}, {Suntzeff}, \& {Taddia}}]{wya21}
{Wyatt}, S.~D., {Sand}, D.~J., {Hsiao}, E.~Y., {et~al.} 2021, \apj, 914, 57,
  \dodoi{10.3847/1538-4357/abf7c3}

\bibitem[{{Wygoda} {et~al.}(2019){Wygoda}, {Elbaz}, \& {Katz}}]{wyg19}
{Wygoda}, N., {Elbaz}, Y., \& {Katz}, B. 2019, \mnras, 484, 3951,
  \dodoi{10.1093/mnras/stz146}

\bibitem[{{Zhang} {et~al.}(2015){Zhang}, {Wang}, {Chen}, {Zhang}, {Zhou}, {Li},
  {Liu}, {Mo}, {Zhang}, {Yao}, {Zhao}, {Zhou}, {Nie}, {Huang}, {Jiang}, {Ma},
  {Wang}, {Wu}, {Zhou}, {Zou}, \& {Wang}}]{zhangtm15}
{Zhang}, T.-M., {Wang}, X.-F., {Chen}, J.-C., {et~al.} 2015, Research in
  Astronomy and Astrophysics, 15, 215, \dodoi{10.1088/1674-4527/15/2/006}

\bibitem[{{Zhao} {et~al.}(2015){Zhao}, {Wang}, {Maeda}, {Sai}, {Zhang},
  {Zhang}, {Huang}, {Rui}, {Zhou}, \& {Mo}}]{zhao15}
{Zhao}, X., {Wang}, X., {Maeda}, K., {et~al.} 2015, \apjs, 220, 20,
  \dodoi{10.1088/0067-0049/220/1/20}

\bibitem[{{Zhao} {et~al.}(2016){Zhao}, {Maeda}, {Wang}, {Wang}, {Sai}, {Zhang},
  {Zhang}, {Huang}, \& {Rui}}]{zhao16}
{Zhao}, X., {Maeda}, K., {Wang}, X., {et~al.} 2016, \apj, 826, 211,
  \dodoi{10.3847/0004-637X/826/2/211}

\end{thebibliography}

\clearpage

\begin{longrotatetable}
\begin{deluxetable}{lcrccccccccr}
\centerwidetable
\tabletypesize{\footnotesize}
\tablecaption{Optical photometric observations of SN2012ij \label{table12ij}}
\tablehead{
\colhead{UT Date} & \colhead{MJD} & 
\colhead{Phase\tablenotemark{a}} & \colhead{$B$} & 
\colhead{$V$} & \colhead{$R$} & 
\colhead{$I$} & \colhead{$u$} & 
\colhead{$g$} & \colhead{$r$} & \colhead{$i$} &
\colhead{Telescope}}
\startdata
2013 Jan. 4& 56295.84& $-$6.5& 16.57(0.01)&   16.06(0.02)&   16.01(0.04)&    16.26(0.02)&  \nodata&  \nodata&  \nodata&  \nodata&  TNT\\
2013 Jan. 5& 56296.89& $-$5.4& 16.30(0.03)&   15.93(0.03)&   15.72(0.02)&    15.86(0.02)&  \nodata&  \nodata&  \nodata&   \nodata& TNT\\
2013 Jan. 6& 56297.83& $-$4.5& 16.08(0.01)&   15.76(0.02)&   15.64(0.03)&    15.64(0.02)&  \nodata&  \nodata&  \nodata&   \nodata& TNT\\
2013 Jan. 6& 56298.32& $-$4.0& 15.90(0.01)& 15.74(0.01)&   \nodata&  \nodata& 16.80(0.04)& 15.75(0.01)& 15.74(0.01)& 15.88(0.01)& Swope\\
2013 Jan. 7& 56298.86& $-$3.4& 16.00(0.01)&   15.66(0.03)&   15.49(0.03)&    15.55(0.02)&  \nodata&  \nodata&  \nodata&   \nodata& TNT\\
2013 Jan. 7& 56299.35& $-$3.0& 15.79(0.01)& 15.66(0.01)&   \nodata&  \nodata& 16.74(0.04)& 15.62(0.01)& 15.63(0.02)& 15.84(0.02)& Swope\\
2013 Jan. 8& 56300.33& $-$2.0& 15.62(0.02)& 15.46(0.03)&   \nodata&    \nodata&  \nodata& 15.49(0.02)& 15.47(0.02)& 15.69(0.02)& Swope\\
2013 Jan. 9& 56300.83& $-$1.5& 15.77(0.01)&   15.38(0.02)&   15.48(0.02)&    15.50(0.02)&  \nodata&  \nodata&  \nodata&   \nodata& TNT\\
2013 Jan. 9& 56301.34& $-$1.0& 15.56(0.02)& 15.44(0.03)&    \nodata&  \nodata& 16.69(0.04)& 15.42(0.02)& 15.41(0.02)& 15.70(0.02)& Swope\\
2013 Jan. 10& 56301.90 & $-$0.4& 15.67(0.01)&   15.26(0.02)&   15.37(0.02)&    15.46(0.01)&  \nodata&  \nodata&  \nodata&   \nodata& TNT\\
2013 Jan. 12& 56303.84& 1.5& 15.74(0.01)&   15.26(0.02)&   15.30(0.04)&    15.47(0.03)&  \nodata&  \nodata&  \nodata&   \nodata& TNT\\
2013 Jan. 12& 56304.37& 2.0& 15.76(0.03)& 15.39(0.03)&   \nodata&    \nodata&  \nodata& 15.49(0.02)& 15.34(0.02)& 15.74(0.02)& Swope\\
2013 Jan. 13& 56304.88& 2.6& 15.87(0.04)&   15.29(0.05)&   15.40(0.05)&    15.49(0.08)&  \nodata&  \nodata&  \nodata&   \nodata& TNT\\
2013 Jan. 13& 56305.33& 3.0& 15.74(0.02)& 15.37(0.02)&    \nodata&  \nodata& 17.08(0.04)& 15.51(0.02)& 15.35(0.02)& 15.77(0.02)& Swope\\
2013 Jan. 14& 56305.83& 3.5& 15.94(0.02)&   15.34(0.03)&   15.34(0.03)&    15.54(0.02)&  \nodata&  \nodata&  \nodata&   \nodata& TNT\\
2013 Jan. 14& 56306.34& 4.0& 15.89(0.02)& 15.42(0.02)&   \nodata&  \nodata& 17.29(0.04)& 15.58(0.02)& 15.39(0.02)& 15.81(0.02)& Swope\\
2013 Jan. 15& 56307.33& 5.0& 16.01(0.03)& 15.45(0.02)&  \nodata&  \nodata& 17.33(0.03)& 15.66(0.02)& 15.42(0.02)& 15.88(0.02)& Swope\\
2013 Jan. 16& 56307.75& 5.4& 16.14(0.02)&   15.47(0.01)&   15.45(0.02)&    15.70(0.02)&  \nodata&  \nodata&  \nodata&   \nodata& TNT\\
2013 Jan. 16& 56308.33& 6.0& 16.21(0.01)& 15.59(0.01)&   \nodata&  \nodata& 17.51(0.02)& 15.83(0.01)& 15.54(0.02)& 15.99(0.01)& Swope\\
2013 Jan. 17& 56309.35& 7.0& 16.30(0.03)& 15.57(0.02)&   \nodata&    \nodata&  \nodata& 15.91(0.02)& 15.56(0.02)& 16.01(0.02)& Swope\\
2013 Jan. 18& 56309.88& 7.6& 16.55(0.02)&   15.66(0.02)&   15.55(0.04)&    15.96(0.02)&  \nodata&  \nodata&  \nodata&   \nodata& TNT\\
2013 Jan. 18& 56310.33& 8.0& 16.43(0.03)& 15.68(0.02)& \nodata&  \nodata& 17.96(0.02)& 16.05(0.02)& 15.60(0.02)& 16.03(0.02)& Swope\\
2013 Jan. 19& 56311.33& 9.0& 16.64(0.03)& 15.80(0.02)&  \nodata&  \nodata& 18.16(0.04)& 16.19(0.02)& 15.66(0.02)& 16.08(0.02)& Swope\\
2013 Jan. 20& 56312.32& 10.0& 16.93(0.04)& 15.89(0.03)&  \nodata&  \nodata& 18.45(0.09)& 16.34(0.03)& 15.78(0.02)& 16.08(0.03)& Swope\\
2013 Jan. 22& 56314.33& 12.0& 17.18(0.03)& 16.14(0.03)&  \nodata&  \nodata& 18.66(0.02)& 16.67(0.02)& 15.89(0.02)& 16.13(0.02)& Swope\\
2013 Jan. 23& 56315.36& 13.0& 17.33(0.03)& 16.24(0.04)&  \nodata&  \nodata& 18.78(0.02)& 16.80(0.02)& 15.93(0.02)& 16.11(0.02)& Swope\\
2013 Jan. 24& 56315.80& 13.5& 17.45(0.02)&   16.24(0.04)&   15.93(0.01)&    15.97(0.02)&  \nodata&  \nodata&  \nodata&   \nodata& TNT\\
2013 Jan. 24& 56316.30& 14.0& 17.48(0.03)& 16.32(0.02)&  \nodata& \nodata& 19.02(0.06)& 16.94(0.02)& 16.02(0.02)& 16.14(0.02)& Swope\\
2013 Jan. 25& 56316.87& 14.6& 17.53(0.02)&   16.39(0.07)&   16.08(0.04)&    16.00(0.05)&  \nodata&  \nodata&  \nodata&   \nodata& TNT\\
2013 Jan. 25& 56317.29& 15.0& 17.66(0.03)& 16.46(0.03)&  \nodata& \nodata& 19.10(0.07)& 17.02(0.03)& 16.05(0.02)& 16.18(0.02)& Swope\\
2013 Jan. 26& 56317.85& 15.6& 17.54(0.06)&   16.45(0.05)&   16.02(0.02)&    15.94(0.03)&  \nodata&  \nodata&  \nodata&   \nodata& TNT\\
2013 Jan. 27& 56318.85& 16.6& 18.03(0.03)&   16.64(0.10)&   16.12(0.07)&    15.85(0.04)&  \nodata&  \nodata&  \nodata&   \nodata& TNT\\
2013 Jan. 27& 56319.32& 17.0& 17.86(0.04)& 16.62(0.03)& \nodata&  \nodata& 19.21(0.10)& 17.42(0.04)& 16.20(0.02)& 16.26(0.02)& Swope\\
2013 Jan. 29& 56320.78& 18.5& 18.07(0.04)&   16.78(0.09)&   16.28(0.03)&    15.95(0.02)&  \nodata&  \nodata&  \nodata&   \nodata& TNT\\
2013 Feb. 1& 56324.28& 22.0& 18.19(0.03)& 16.93(0.03)&  \nodata&  \nodata& 19.57(0.09)& 17.60(0.03)& 16.62(0.02)& 16.58(0.02)& Swope\\
2013 Feb. 2& 56324.80& 22.5& 18.24(0.03)&   16.96(0.02)&   16.71(0.02)&    16.14(0.02)&  \nodata&  \nodata&  \nodata&   \nodata& TNT\\
2013 Feb. 2& 56325.27& 23.0& 18.16(0.03)& 17.02(0.02)&   \nodata&  \nodata& 19.45(0.05)& 17.69(0.02)& 16.70(0.02)& 16.67(0.02)& Swope\\
2013 Feb. 5& 56328.29& 26.0& 18.19(0.03)& 17.12(0.02)&  \nodata&  \nodata& 19.51(0.06)& 17.74(0.02)& 16.93(0.02)& 16.89(0.02)& Swope\\
2013 Feb. 6& 56328.80& 26.5& 18.14(0.04)&   17.14(0.02)&   16.97(0.03)&    16.68(0.02)&  \nodata&  \nodata&  \nodata&   \nodata& TNT\\
2013 Feb. 7& 56330.25& 28.0& 18.39(0.02)& 17.34(0.02)&  \nodata&  \nodata& 19.63(0.17)& 18.04(0.03)& 17.14(0.02)& 17.07(0.01)& Swope\\
2013 Feb. 14& 56337.39& 35.0& 18.65(0.03)& 17.67(0.03)&  \nodata&    \nodata&  \nodata& 18.28(0.03)& 17.48(0.02)& 17.48(0.02)& Swope\\
2013 Feb. 16& 56338.66& 36.4& 18.87(0.07)&   17.78(0.05)&   17.44(0.02)&    17.46(0.02)&  \nodata&  \nodata&  \nodata&   \nodata& TNT\\
2013 Feb. 21& 56344.25& 42.0& 18.77(0.03)& 17.85(0.03)&   \nodata&    \nodata&  \nodata& 18.47(0.03)& 17.85(0.02)& 17.76(0.02)& Swope\\
2013 Mar. 2& 56353.32& 51.0& 19.06(0.04)& 18.13(0.03)&   \nodata&  \nodata& 20.55(0.26)& 18.67(0.03)& 18.15(0.02)& 18.23(0.02)& Swope\\
2013 Mar. 9& 56360.30& 58.0&  \nodata&  \nodata&   \nodata&    \nodata&  \nodata&  \nodata& 18.51(0.02)&  \nodata& Swope\\
2013 Mar. 16& 56367.26& 65.0& 19.31(0.03)& 18.57(0.03)&    \nodata& \nodata&   \nodata& 18.74(0.02)& 18.78(0.03)& 18.71(0.03)& Swope\\
\enddata
\tablenotetext{a}{Relative to the epoch of $B$-band maximum (MJD = 56302.3) in the frame of the observer.}
\end{deluxetable}
\end{longrotatetable}

\clearpage

\begin{deluxetable}{lcrlcr}
\tabletypesize{\small}
\tablecaption{\label{table12ijspec}Journal of spectroscopic observations of SN 2012ij}
\tablehead{\colhead{UT Date} & \colhead{MJD} & 
\colhead{Phase\tablenotemark{a}} & \colhead{Range(\AA)} & 
\colhead{Resolution(\AA)\tablenotemark{b}} & \colhead{Instrument}}
\startdata
2013 Jan. 5& 56297.0& $-$5.3& 3500-8800& 3& LJT YFOSC\\
2013 Jan. 5& 56297.8& $-$4.5& 3500-8500& 5& XLT BFOSC\\
2013 Jan. 6& 56298.3& $-$4.0& 8000-25800& 4-16&\textit{Magellan Baade} Telescope FIRE\\
2013 Jan. 7& 56299.3& $-$3.0& 3500-9500& 2& du Pont WF\\
2013 Jan. 8& 56300.8& $-$1.5& 3400-8700& 5& XLT BFOSC\\
2013 Jan. 10& 56302.3& 0.0& 3500-9500& 2& du Pont WF\\
2013 Jan. 12& 56304.0& 1.7& 3500-8800& 3& LJT YFOSC\\
2013 Jan. 20& 56312.0& 9.7& 3500-8800& 3& LJT YFOSC\\
2013 Feb. 1& 56324.8& 22.5& 4200-8800& 5& XLT BFOSC\\
2013 Feb. 2& 56325.2& 22.9& 8000-25800& 4-16&\textit{Magellan Baade} Telescope FIRE\\
2013 Feb. 3& 56327.1& 24.8& 3500-9000& 3& NOT ALFOSC\\
2013 Feb. 5& 56328.3& 26.0& 8000-25800& 4-16&\textit{Magellan Baade} Telescope FIRE\\
2013 Feb. 15& 56338.7& 36.4& 3700-8700& 5& XLT BFOSC\\
2013 Feb. 18& 56341.3& 39.0& 3300-9500& 3& du Pont BC\\
\enddata
\tablenotetext{a}{Relative to the epoch of $B$-band maximum (MJD = 56302.3) in the frame of the observer.}
\tablenotetext{b}{Approximate spectral resolution (FWHM intensity).}
\end{deluxetable}

\begin{deluxetable}{lcrcc}
\tabletypesize{\small}
\tablecaption{\label{table12ijunf}Unfiltered photometric observations of SN 2012ij by 0.6-m Schmidt Telescope}
\tablehead{\colhead{UT Date} & \colhead{MJD}  & \colhead{Phase\tablenotemark{a}} & \colhead{Magnitude} & \colhead{Error}}
\startdata
2012 Dec. 30& 56290.80 & $-$11.5 & 17.97& 0.24\\
2013 Jan. 1& 56292.80 & $-$9.5 & 17.03& 0.20\\
2013 Jan. 3& 56294.50 & $-$7.8 & 16.20& 0.18\\
2013 Jan. 24& 56315.50 & 13.2 & 15.75& 0.19\\
2013 Jan. 24& 56315.50 & 13.2 & 15.84& 0.19\\
2013 Jan. 27& 56318.50 & 16.2 & 16.17& 0.21\\
\enddata
\tablenotetext{a}{Relative to the epoch of $B$-band maximum (MJD = 56302.3) in the frame of the observer.}
\end{deluxetable}

\begin{deluxetable}{lll}
\tabletypesize{\small}
\tablecaption{\label{tableof2obj}Photometric and spectroscopic parameters of SN 2012ij}
\tablehead{\colhead{SN 2012ij} & \colhead{}& \colhead{}}
\startdata
RA(J2000)     & 11$^{\rm{h}}$40$^{\rm{m}}$15$^{\rm{s}}$.84 \\
Dec(J2000)    & +17$^{\circ}$27$\arcmin$22$\arcsec$.2 \\
Date of $B_{\rm{max}}$(MJD)      & 56302.3 $\pm$ 0.3    \\
$E(B-V)_{\rm{MW}}$   & 0.0235  \\ 
${\Delta}m_{15}(B)$  & 1.86 $\pm$ 0.05 (mag)  \\
$s_{BV}$             & 0.53 $\pm$ 0.06    \\
Redshift             & 0.011 \\
Distance modulus     & 33.54 $\pm$ 0.15 (mag)      \\ 
$m_{B,\rm{max}}$     & 15.67 $\pm$ 0.01 (mag)     \\ 
$M_{B,\rm{max}}$     & $-$17.95 $\pm$ 0.15 (mag)     \\ 
pEW(Si~{\sc ii} $\lambda$5972) &  54 $\pm$ 3 (\AA) \\
pEW(Si~{\sc ii} $\lambda$6355) &  109 $\pm$ 4 (\AA)\\
Host mophology       & S0/a  \\
\enddata
\end{deluxetable}

\begin{deluxetable}{lllllll}
\tabletypesize{\small}
\tablecaption{\label{tableOthers}Photometric properties of some SNe Ia in comparison}
\tablehead{\colhead{name} & \colhead{$s_{BV}$}  & \colhead{$M_{B, \rm{max}}$(mag)} & \colhead{${\Delta}m_{15}(B)$(mag)} & \colhead{host morphology}& \colhead{Ref.\tablenotemark{a}}}
\startdata
SN1986G   & 0.54 & $-$17.76 $\pm$ 0.32 & 1.69 & S0      &4\\
SN1991bg  & 0.34 & $-$16.85 $\pm$ 0.34 & 1.92 & E1      &5,6\\
SN1997cn  & 0.40 & $-$17.17 $\pm$ 0.2  & 1.84 & E       &7\\
SN1998de  & 0.36 & $-$16.74 $\pm$ 0.19 & 1.91 & S0/a    &9,10\\
SN1999by  & 0.43 & $-$17.17 $\pm$ 0.26 & 1.88 & Sb      &8,10,11\\
SN1999da  & 0.37 & $-$16.68 $\pm$ 0.22 & 2.11 & E       &10,11\\
SN2002cf  & 0.40 & $-$17.46            & 1.78 & E/S0    &10\\
SN2002dl  & 0.55 & $-$18.33            & 1.82 & Sbc     &11\\
SN2002es  & 0.60 & $-$17.78 $\pm$ 0.12 & 1.28 & S0      &11\\
SN2002fb  & 0.57 & $-$17.24            & 2.02 & E       &11\\
SN2003gs  & 0.45 & $-$17.94 $\pm$ 0.29 & 1.88 & SB0     &10,11,12\\ 
SN2003Y   & 0.40 & $-$16.7             & 1.95 & S0      &10,11\\
SN2005M\tablenotemark{a} & 1.21& $-$19.5  & 0.799 & S? &1,19\\
SN2005bl  & 0.39 & $-$17.24 $\pm$ 0.34 & 1.93 & E       &1\\
SN2005ke  & 0.42 & $-$17.0 $\pm$ 0.2   & 1.85 & Sa      &1\\
SN2006em  & 0.31 & $-$16.8             & 1.80 & E       &10\\
SN2006gt  & 0.56 & $-$18.2             & 1.99 & E       &1\\
SN2006H   & 0.60 & $-$16.6             & 2.28 & S0      &13\\
SN2006hb  & 0.66 & $-$18.3             & 1.78 & E       &1\\
SN2007ba  & 0.55 & $-$17.8             & 1.91 & S0/a    &1\\
SN2008R   & 0.60 & $-$18.5             & 1.78 & SA0     &1\\
SN2009dc\tablenotemark{b} & 1.29 & $-$20.22 $\pm$ 0.3  & 0.71 & S0 &2\\
SN2015bp  & 0.63 & $-$17.9             & 1.62 & E       &3\\
SN2017fzw & 0.63 & $-$17.4             & 1.88 & SB0     &3\\
\enddata
\tablerefs{1 - \citet{burns14}; 2 - \citet{ash20}; 3 - \citet{brown14}; 4 - \citet{phi87}; 5 - \citet{lei93}; 6 - \citet{tur96}; 7 - \citet{tur98}; 8 - \citet{gar04}; 9 - \citet{mod01}; 10 - \citet{silver12}; 11 - \citet{gane10}; 12 - \citet{kri09}; 13 - \citet{hic09}.
Some data are obtained through the Open Supernova Catalog \citep{snespace}.}
\tablenotetext{a}{91T-like SN.}
\tablenotetext{b}{super-Chandrasekhar SN.}
\end{deluxetable}

\clearpage

\begin{deluxetable}{lllll}
\tabletypesize{\small}
\tablecaption{\label{table:trans} Main parameters of supernovae in comparison}
\tablehead{\colhead{Object} & \colhead{$s_{BV}$} & \colhead{t$^{i-B}_{\rm{max}}$(days)}  & \colhead{pEW(Ti~{\sc ii})(\AA)}  & \colhead{$\mathcal{\overline{F}}_{i2}$}}
\startdata
SN2007ax	&0.36(0.041)	&4.17(0.21)	    &221     	&0.22(0.01)\\
SN2005ke	&0.42(0.040)	&2.8(0.205)	    &264	    &0.27(0.01)\\
SN2011jq	&0.5(0.041)	    &0.89(0.178)	&\nodata	&0.32(0.03)\\
SN2012ij	&0.53(0.04)	    &$-$0.06(0.16)	&196	    &0.34(0.01)\\
SN2007ba	&0.54(0.041)	&0.48(0.23)	    &232    	&0.39(0.04)\\
SN2006gt	&0.56(0.04)	    &$-$0.81(0.24)	&144    	&0.38(0.01)\\
SN2007on	&0.57(0.04)	    &$-$1.99(0.22)	&91     	&0.40(0.01)\\
iPTF13ebh	&0.61(0.04)	    &$-$1.3(0.11)	    &\nodata	&0.40(0.04)\\
SN2004gs	&0.70(0.04)	    &$-$1.9(0.40)	    &120	    &0.45(0.01)\\
SN2006ob	&0.72(0.04)	    &$-$1.2(0.41)	    &36     	&0.45(0.01)\\
SN2006bh	&0.80(0.04)	    &$-$3.59(0.23)	&\nodata  	&0.50(0.01)\\
SN2004ey    &1.01(0.04)     &$-$2.98(0.16)    &70         &0.50(0.01)\\
\enddata
\tablerefs{The $s_{BV}$ and t$^{i-B}_{\rm{max}}$ are obtained from \citet{ash20}}
\end{deluxetable}

\begin{deluxetable}{llllllll}
\tabletypesize{\small}
\tablecaption{\label{table:models}Main parameters of models in comparison}
\tablehead{\colhead{Model} & \colhead{$M_{\rm{tot}}$($M_{\odot}$)} & \colhead{$M(^{56}\rm{Ni})$($M_{\odot}$)}  & \colhead{$\rho_{\rm{tr}}$ (g cm$^{-3}$)\tablenotemark{a}}  & \colhead{t$_{\rm{rise}}(B)$(day)}  & \colhead{$M_{B}$(mag)}  & \colhead{$\Delta m_{15}(B)$(mag)} & \colhead{ref.}}
\startdata
SCH2p0 & 0.90 & 0.12 & \nodata& 14.6 & $-$17.27 & 1.64 & 1 \\
DDT08  & 1.4 & 0.095& 8.0(6) & 14.38& $-$16.76 & 1.99 & 2 \\
DDT14 & 1.4  & 0.154 & 14.0(6) & 14.86 & $-$17.50 & 1.81 & 2\\ 
\enddata
\tablenotetext{a}{Numbers in parentheses correspond to powers of 10.}
\tablerefs{1 - \citet{blond17}; 2 - \citet{hoe17}}
\end{deluxetable}

\clearpage

\end{document}